\documentclass[preprintnumbers,showpacs,amsmath,amssymb,floatfix,9pt,prd,onecolumn,
superscriptaddress,nofootinbib]{revtex4}
\usepackage{latexsym}
\usepackage{epsfig}
\usepackage{amssymb}

\begin{document}

\title{Noncommutative Wormholes in $f(R)$ Gravity with Lorentzian Distribution}

\author{ Farook Rahaman }
\email{rahaman@iucaa.ernet.in } \affiliation{Department of Mathematics, Jadavpur University, Kolkata 700 032, West Bengal, India}
\author{Ayan Banerjee}
\email{  ayan_7575@yahoo.co.in} \affiliation{Department of Mathematics, Jadavpur University, Kolkata 700 032, West Bengal, India}

\author{Mubasher Jamil}
\email{mjamil@camp.nust.edu.pk} \affiliation{School of Natural Sciences, National University of Sciences and Technology (NUST),
H-12, Islamabad, Pakistan}

\author{Anil Kumar Yadav}
\email{abanilyadav@yahoo.co.in} \affiliation{Department of Physics, Anand Engineering College, Keetham, Agra 282 007, India}

\author{Humaira Idris}
\email{humaira.idris@gmail.com} \affiliation{School of Natural
Sciences, National University of Sciences and Technology (NUST),
H-12, Islamabad, Pakistan}

\begin{abstract}
{\bf Abstract:}   In this paper, we derive some new exact solutions
of static wormholes in $f(R)$ gravity supported  by the matter
possesses Lorentizian density distribution of a particle-like
gravitational source.  We derive the wormhole's solutions in two
possible schemes for a given Lorentzian distribution: assuming an
astrophysically viable $F(R)$ function such as a power-law  form and
discuss several solutions corresponding to different values of the
exponent  ( here $F =\frac{df}{dR}$ ). In the second scheme, we
consider particular form of two shape functions and have
reconstructed f(R) in both cases. We have discussed all the
solutions with graphical point of view.

\end{abstract}


\maketitle

\section{Introduction}

The geometry of a static and stationary spherically  symmetric
wormhole consists of a two-mouthed tunnel (referred as a tube,
throat, or handle in the literature). This tube like structure is a
multiply connected spacetime which can join two asymptotically flat
regions of the same spacetime or two different spacetimes. The
possibility of two way travel in time becomes theoretically possible
when the two regions belong to the same spacetime, such phenomenon
is time machine. The wormhole spacetime has been investigated in the
literature in various schemes such as a non-static axially symmetric
wormhole \cite{teo}, wormholes with cylindrical symmetry \cite{ger},
wormholes supported by a cosmological constant \cite{lemos1},
thin-shell wormholes \cite{lemos}, and electrically charged static
wormholes \cite{kim} (For a review see \cite{lobo}). Morris and
Thorne \cite{morris} suggested the idea of a traversable wormhole
suitable for travel by humans for interstellar journey and beyond.
Later on wormholes were investigated in the cosmological context and
theorized to be enlarged through a mechanism similar to cosmological
inflation \cite{roman}. Wormholes present esoteric properties such
as a violation of the Hawking chronology protection conjecture and
give faster-than-light scenarios, breakdown of causality
\cite{hawking}. The matter energy supporting the exotic geometry
violates the standard energy conditions and hence termed 'exotic'.
However this exotic behavior can be avoided by studying wormholes in
extended theories of gravity such as $f(R)$ gravity and threading
wormholes with normal matter. The violation of energy conditions can
be avoided on account of curvature effects arising near the
wormhole's throat.

In an earlier work by Lobo et al \cite{lobo1},  the authors
investigated the geometry and stability of stationary and static
wormholes in $f(R)$ gravity. They obtained the wormhole's solutions
by assuming various forms of equations of state and viable shape
functions. They showed that in their model,  the energy conditions
 are satisfied in the desired range of radial
coordinate.  We here perform a similar analysis of \cite{lobo1}
however, taking into account a Lorentizian density distribution of a
point gravitational source \cite{hamid}.  We derive the wormhole's
solutions in two possible schemes for a given Lorentzian
distribution: assuming an astrophysically viable $F(R)$ function
such as a power-law form :$F(R) = a R^m$ and have discussed several
solutions corresponding to different values of the exponential
parameter. In the second scheme, for the specific choice of the
shape functions we have reconstructed $f(R)$. We also check the
energy conditions for both schemes.

Our plan  of work  is as follows: In Sec.II, we write down the field
equations of a Morris-Thorne wormhole in $f(R)$ gravity supported by
anisotropic matter.  In Sec.III, we solve the field equations by
assuming a power-law form: $F(R) = a R^m$ and discussed several
solutions corresponding to different values of m. In Sec.IV, we
solve the same field equations by inserting specific forms of shape
functions and have reconstructed $f(R)$ in all cases and we conclude
in Sec. IV.

\section{Field equations  in $F(R)$ gravity}

The metric  describing a static spherically symmetric wormhole spacetime is given by
\begin{equation}
ds^2=  - e^{2\Phi(r)} dt^2+ \frac{ dr^2}{1-\frac{b(r)}{r}}+r^2 (d\theta^2+\sin^2\theta d\phi^2). \label{Eq3}
\end{equation}
Here, $\Phi(r)$ is a gravitational redshift function  and $b(r)$ is
the shape function. The radial co-ordinate r, decreases from infinity
to a minimum value $r_0$, and then increases from $r_0$ to infinity. The
minimum value of $r_0$, represents the location of the wormhole throat
where $b(r_0) =r_0$, satisfying the flaringout condition $b-b^{\prime}r/b^2 > 0$
and $b^{\prime}(r_0)< 1$, that are imposed to have wormhole solution.

   For our wormhole  in F(R)    gravity,
we assume that the matter content of the wormhole is anisotropic fluid
source whose energy-momentum tensor is given by \cite{lobo}
\begin{equation}
T_\nu^\mu=  ( \rho + p_r)u^{\mu}u_{\nu} - p_r g^{\mu}_{\nu}+ (p_t -p_r )\eta^{\mu}\eta_{\nu}, \label{eq:emten}
\end{equation}
with $u^{\mu}u_{\mu} = - \eta^{\mu}\eta_{\mu} = 1, $ and $u^\mu
\eta_\mu=0.$  Here the vector $u^\mu$ is the fluid 4-velocity and
$\eta^\mu$ is a space-like vector orthogonal to $u^\mu$.

Following Lobo et al's   \cite{lobo1}, we have the following
gravitational field equations in $f(R)$ gravity   as
\begin{eqnarray}
\rho(r) &=& \frac{F b'}{r^2},\label{r}\\
p_r(r)&=& -\frac{F b}{r^3} +\frac{F'}{2r^2}(b'r-b)-F''
\Big(1-\frac{b}{r}\Big),\label{r1}\\
p_t(r)&=& -\frac{F'}{r}\Big(1-\frac{b}{r}\Big)
-\frac{F}{2r^3}(b'r-b).\label{r2}
\end{eqnarray}
Here, prime indicate derivative with respect to r and $\rho$, $p_r$
and $p_t$ are the energy density, radial pressure and tangential
pressure, respectively. These equations are the generic expressions
of the matter threading wormhole with constant redshift function
 which simplifies the calculations of field equations, and provide
interesting exact wormhole solutions,  where $F =\frac{df}{dR}$ and
the curvature scalar, $R$, is given by
\begin{equation}
    R(r) = 2\frac{b'(r)}{r^2}.
   \end{equation}

\section{Wormholes for a given $F(R)$ function}

We take a power-law form
\begin{equation}
F(R) =a R^m.
\end{equation}
Here, $a$ is a constant and m is an integer.

We are trying to  solve the field equations given above  in
non-commutative geometry with Lorentzian distribution. Here we take
the energy density of the static and spherically symmetric smeared
and particle-like gravitational source of the following \cite{hamid}
\begin{equation}
\rho = \frac{M \sqrt{\phi}}{\pi^2(r^2 +\phi)^2}.
\end{equation}
Here,  the mass $M$ could be a diffused centralized object such as a
wormhole and $\theta$ is noncommutative parameter.

Using Eq.(6) in Eq. (7), gives
\begin{equation}
F(R) = a\Big(\frac{2b^{\prime}}{r^2}\Big)^m.
\end{equation}
Substituting Eqs. (8) and (9) in Eq. (3), we obtain the shape function given by
\begin{equation}
b(r)=\int{r^2\Big[\frac{M \sqrt{\phi}}{2^m \pi^2 a(r^2
+\phi)^2}\Big]^\frac{1}{1+m}}+C
\end{equation}
where $C$ is a constant of integration.

To get the exact physical characteristics, we discuss several models resulting for different choices of m.

\subsection{$m=0$}

From Eq.(10), the assumption m=0, gives the shape function of the form
\begin{equation}
b(r) = \frac{M\sqrt{\phi}}{2 a
\pi^2}\Big[\frac{\arctan\Big(\frac{r}{\sqrt{\phi}}\Big)}{\sqrt{\phi}}
-\frac{r}{r^2+\phi} \Big]+C.
\end{equation}
By putting (11) in (6), we get
\begin{equation}
 R(r) = \frac{2M\sqrt{\phi}}{\pi^{2}a(r^2 + \phi)^2}.
\end{equation}
From Eqs. (4) and (5), using the shape function in Eq.(11)
the expressions for pressures (radial and tangential) become
\begin{equation}
 P_{r}(r) = \frac{a}{r^3}\Big[\frac{M\sqrt{\phi}}{2a\pi^{2}}\Big(\frac{r}{r^2 +\phi}-\frac{1}{\sqrt{\phi}}
~ \arctan\Big(\frac{r}{\sqrt{\phi}}\Big)\Big)-C\Big],
\end{equation}
\begin{equation}
 P_{t}(r) = \frac{M\sqrt{\phi}}{4\pi^{2}r^{3}}\Big[\frac{1}{\sqrt{\phi}}~\arctan\Big(\frac{r}{\sqrt{\phi}}\Big) -
\frac{r}{(r^2 + \phi)}-\frac{2r^3}{(r^2 + \phi)^2} +
\frac{2aC\pi^2}{M\sqrt{\phi}}\Big].
\end{equation}

We consider that the Lorentzian distribution of particle-like
gravitational source given in Eq.(8), which is positive for the
noncommutative parameter $\theta > 0$. The shape function given in
Eq.(11) is asymptotically flat because of $b(r)/r \rightarrow 0$ as
$r \rightarrow \infty$ and the redshift function is constant
everywhere. In fig.(1), corresponding to m=0, the throat of the
wormhole is located at r=1.5, where $\mathcal{G}(r)$ = b(r)-r cuts
the r-axis ( upper right in Fig.(1)) and also $\mathcal{G} (r) < 0$,
i.e., b(r)-r$< 0$, which implies that b(r) $<$ r for r$> r_0$,
satisfy the fundamental property of shape function which indicates
in Fig.(1) (upper left). From the Fig.(1), it is also clear that,
for r $> r_0$, $\mathcal{G}(r)$ is decreasing function, therefore
$\mathcal{G^{\prime}}(r) < 0$ and correspondingly  $b^{\prime}(r_0)
< 1$, which satisfies the flare-out condition.

 According to a Fig.(1) (left and right, middle position) the radial pressure $(p_r)$
is negative and transverse pressure $(p_t)$ is positive outside  the
throat of the wormhole for the choice of the parameters M=10, C=1,
a=1 and $\phi= 1$, respectively, which shows that the wormholes
violates the null energy condition (NEC) as well as weak energy
conditions (WEC).

\begin{figure*}[thbp]
\begin{tabular}{rl}
\includegraphics[width=7.5cm]{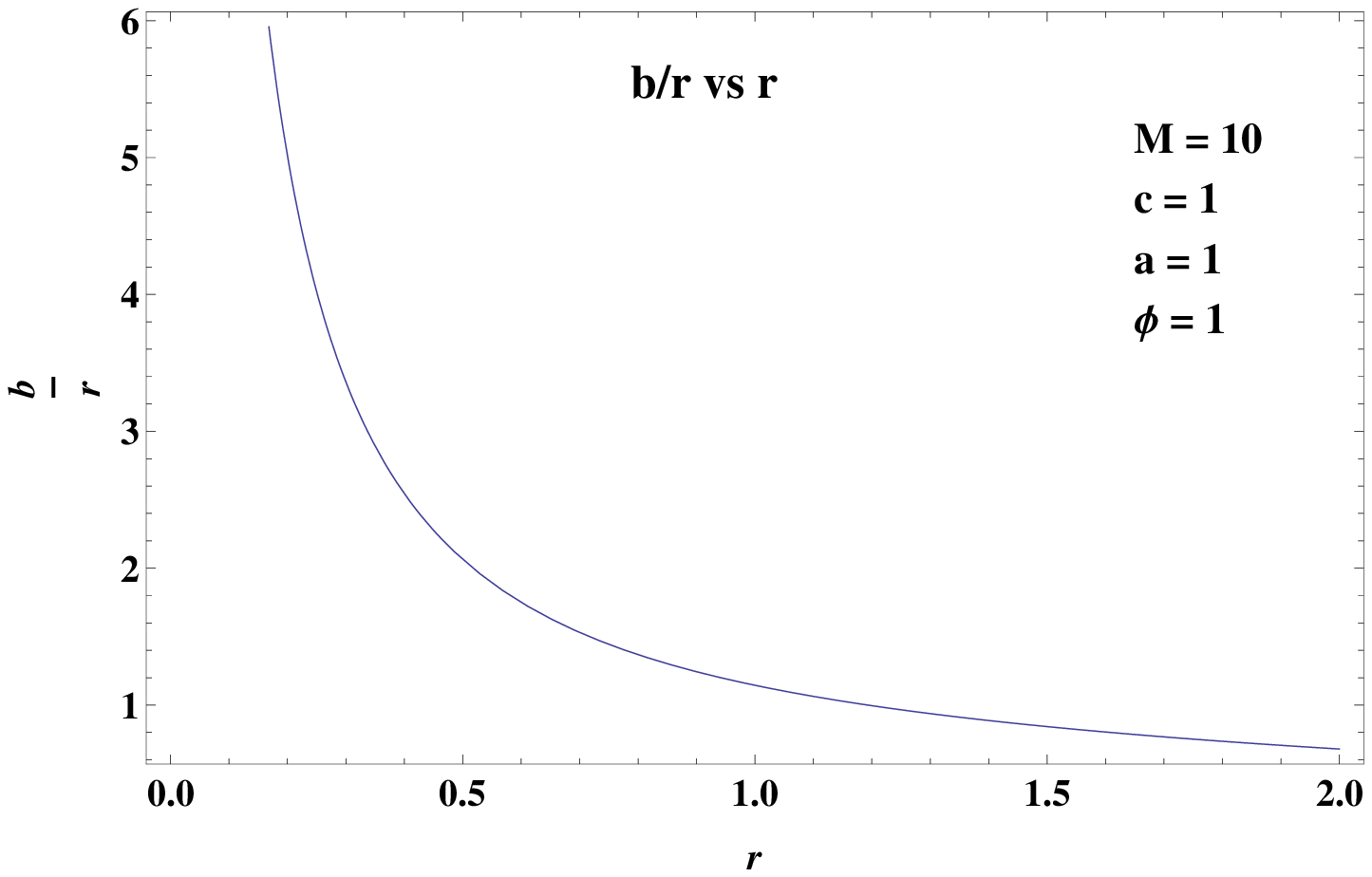}&
\includegraphics[width=7.5cm]{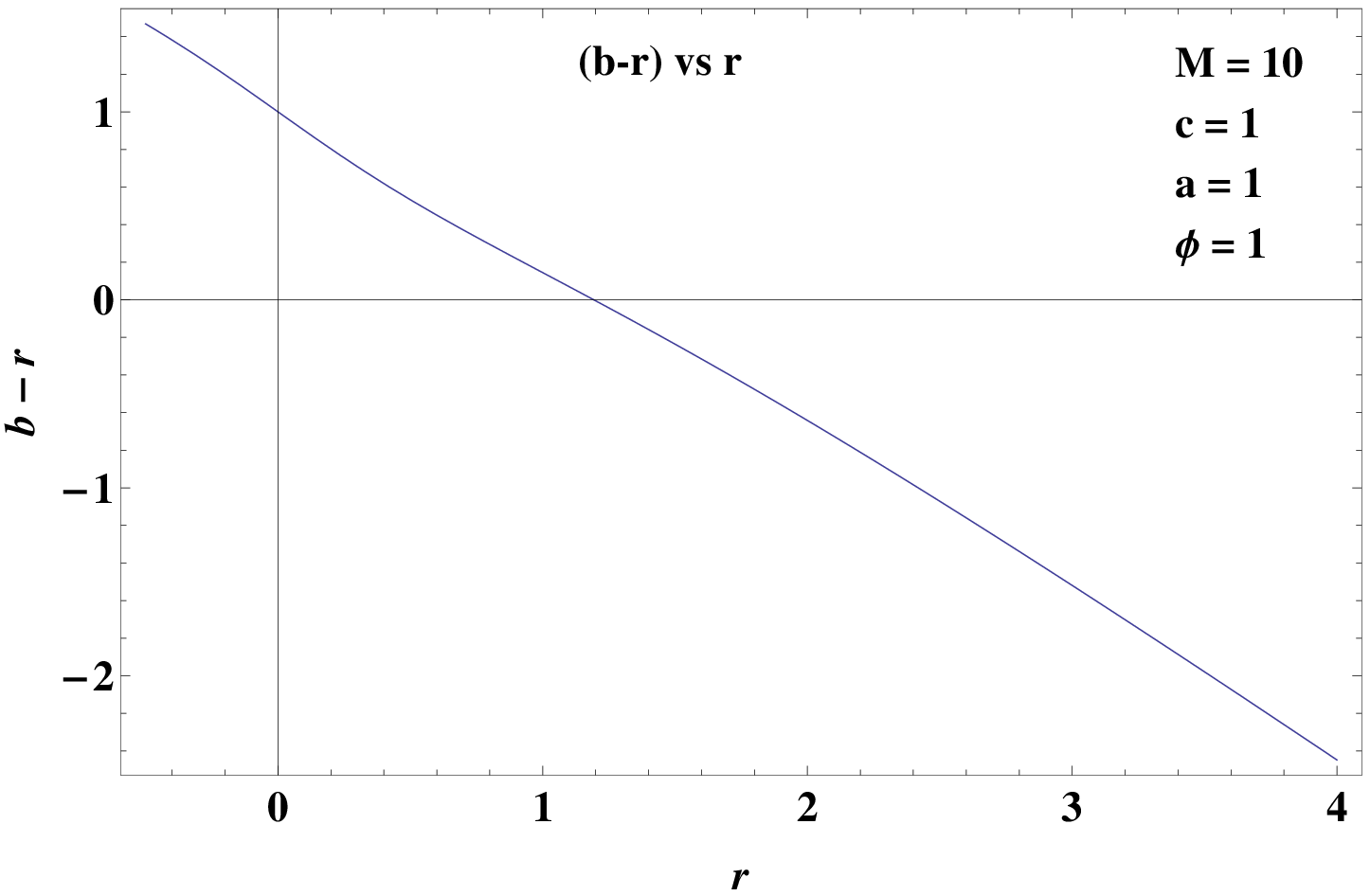} \\
\includegraphics[width=7cm]{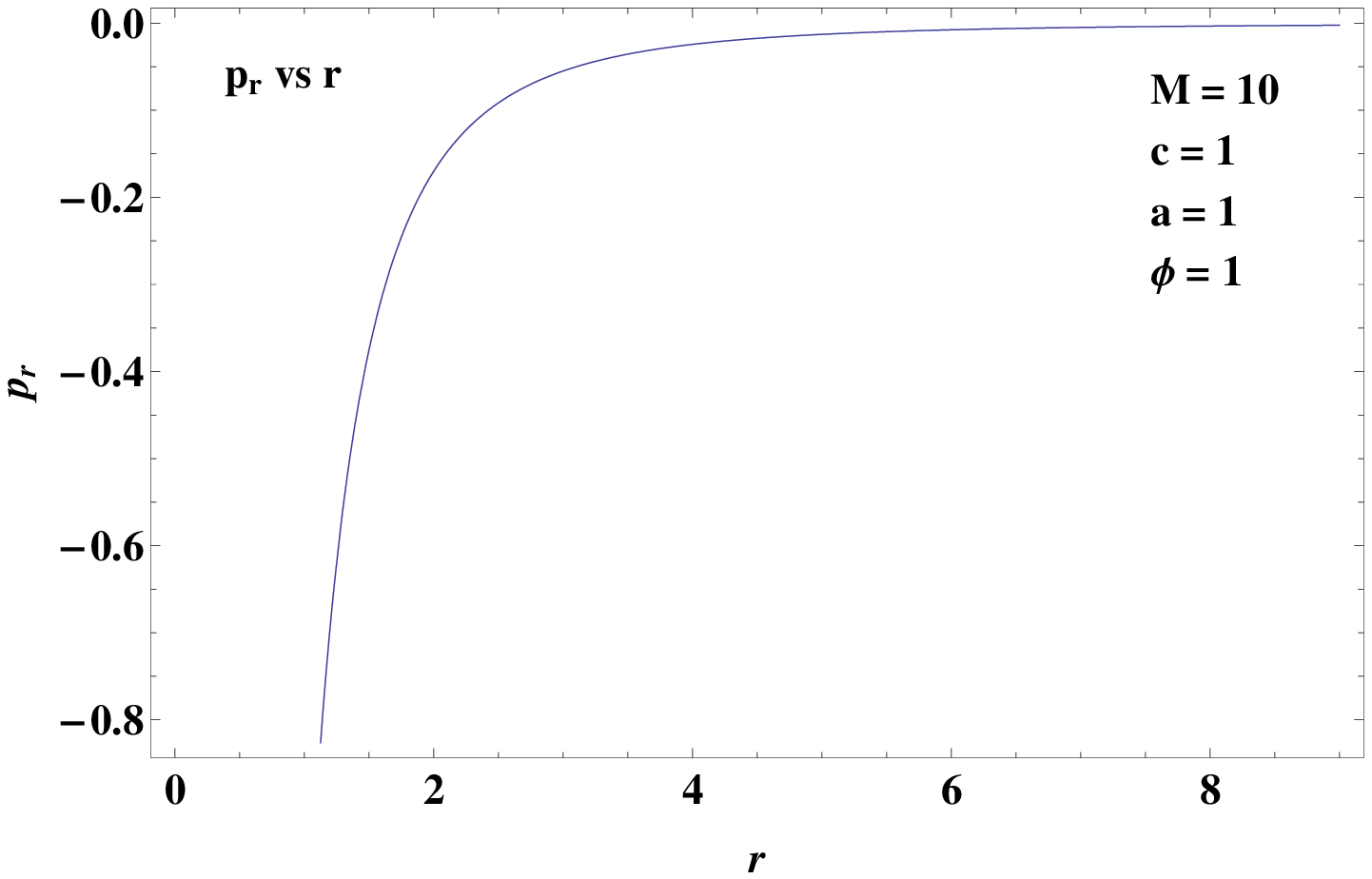}&
\includegraphics[width=7cm]{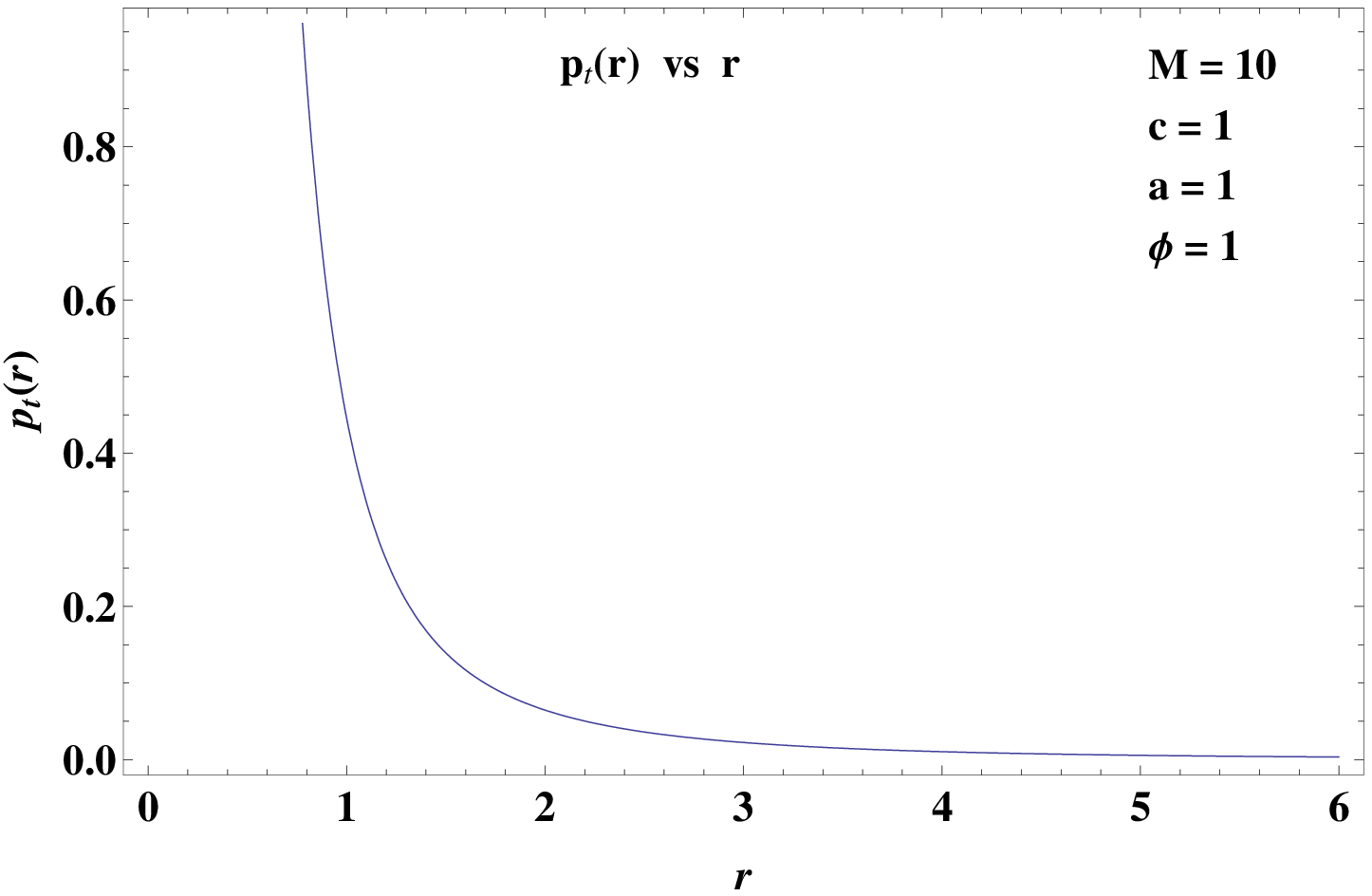} \\
\includegraphics[width=7cm]{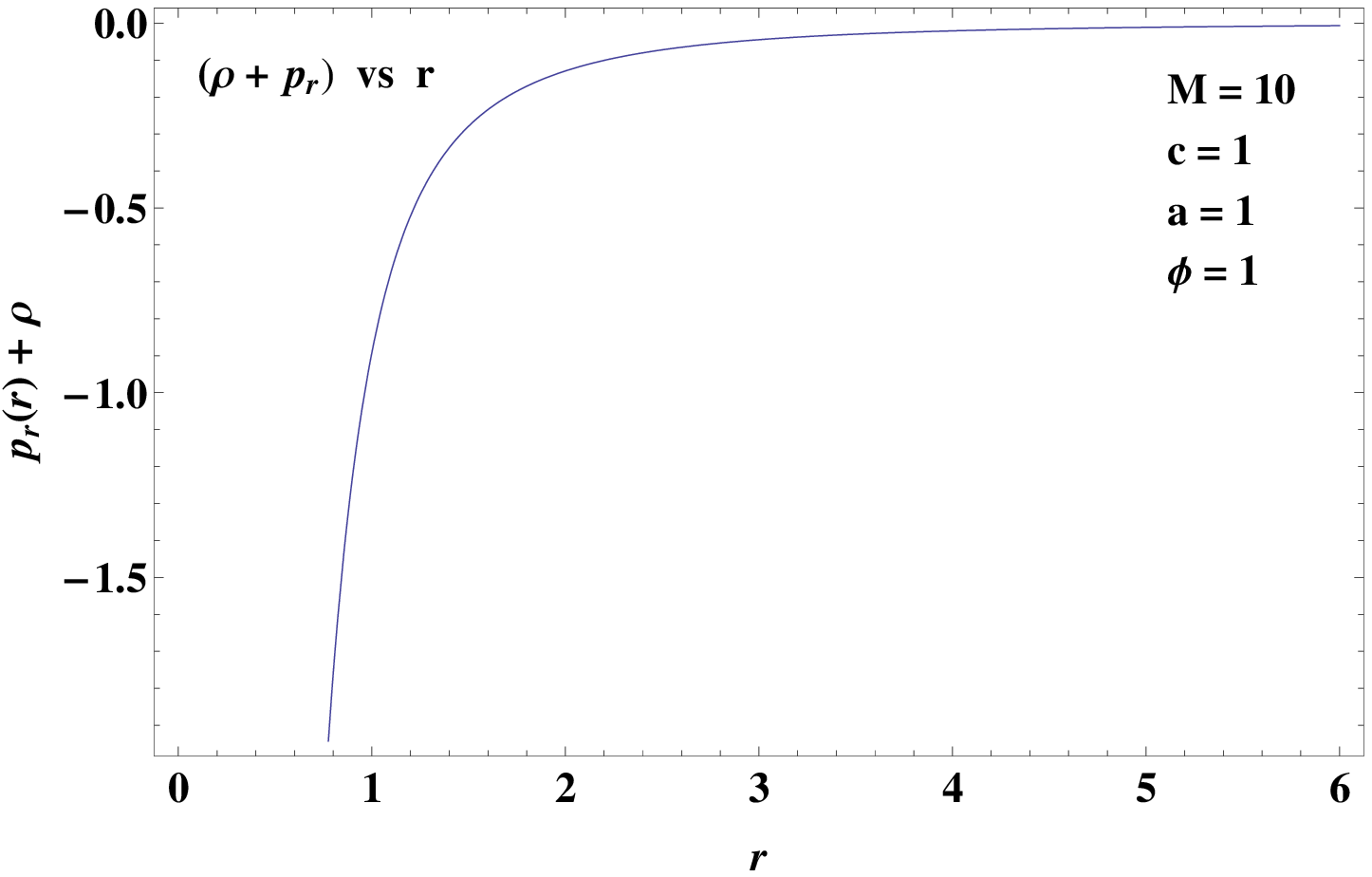} &
\includegraphics[width=7cm]{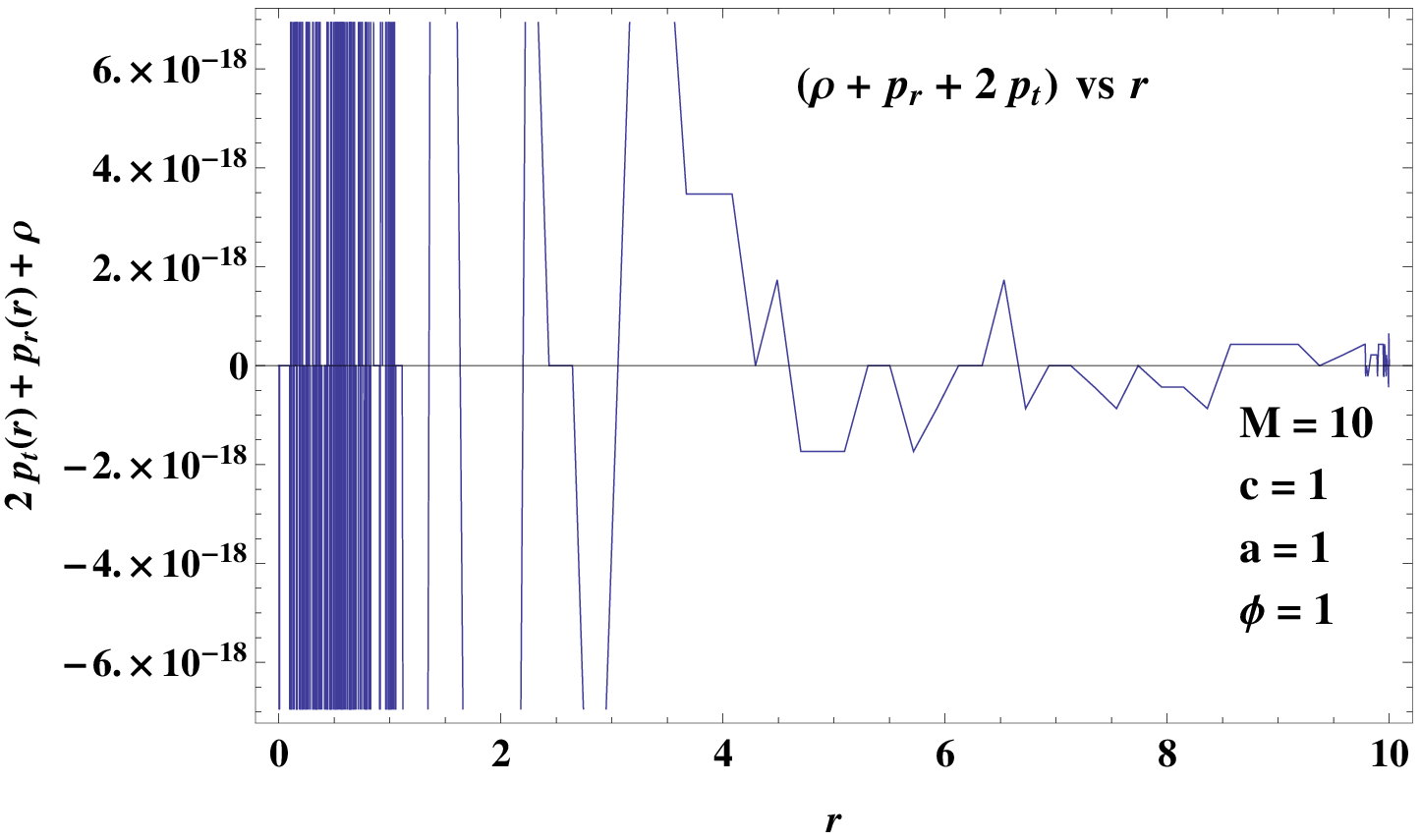} \\
\end{tabular}
\caption{ Graphs for the case $m=0$. }
\end{figure*}

\subsection{$m=1$}

Similarly, from Eq. (10), we get the shape function for m=1
\begin{equation}
b(r) = \sqrt{\frac{M\sqrt{\phi}}{2 a \pi^2}}\Big[r-\sqrt{\phi} ~~
\arctan\Big(\frac{r}{\sqrt{\phi}}\Big)\Big]+C.
\end{equation}
The corresponding value of Ricci scalar becomes
\begin{equation}
 R(r) = \sqrt{\frac{2M\sqrt{\phi}}{a\pi^2(r^2 + \phi)^2}}.
\end{equation}
Moreover, the radial and transverse pressures turn out as
\begin{eqnarray}
P_{r}(r) &=&
\sqrt{\frac{2Ma\sqrt{\phi}}{\pi^2}}\Big[\frac{1}{r^3(r^2 +
\phi)}\Big[\sqrt{\frac{M\sqrt{\phi}}{2a\pi^2}} \Big(\sqrt{\phi}~
\arctan\Big(\frac{r}{\sqrt{\phi}}\Big)-r\Big)-C\Big]\nonumber\\&&-
\frac{1}{r(r^2+\phi)^2}\Big[\sqrt{\frac{M\sqrt{\phi}}{2a\pi^2}}\frac{r^3}{(r^2+\phi)}-
\sqrt{\frac{M\sqrt{\phi}}{2a\pi^2}}\Big(r -
\sqrt{\phi}~\arctan\Big(\frac{r}{\sqrt{\phi}}\Big)\Big)-C\Big]\nonumber\\&&
-\frac{(6r^4+4r^2\phi-2\phi^2)}{(r^2+\phi)^4}\Big[1-
\sqrt{\frac{M\sqrt{\phi}}{2a\pi^2}}\Big(1-\frac{\sqrt{\phi}}{r}~\arctan\Big(\frac{r}{\sqrt{\phi}}\Big)\Big)
-\frac{C}{r}\Big]\Big].
\end{eqnarray}
\begin{eqnarray}
P_{t} (r)&=&
\sqrt{\frac{2Ma\sqrt{\phi}}{\pi^2}}\Big[\frac{2}{(r^2+\phi)^2}\Big[1
- \sqrt{\frac{M\sqrt{\phi}}{2a\pi^2}}
\Big(1-\frac{\sqrt{\phi}}{r}~\arctan\Big(\frac{r}{\sqrt{\phi}}\Big)\Big)-\frac{C}{r}\Big]\nonumber\\&&-
\frac{1}{2r^3(r^2+\phi)}\Big[\sqrt{\frac{M\sqrt{\phi}}{2a\pi^2}}
\frac{r^3}{(r^2+\phi)}-\sqrt{\frac{M\sqrt{\phi}}{2a\pi^2}}\Big(r-\sqrt{\phi}~\arctan\Big(\frac{r}{\sqrt{\phi}}
\Big)\Big)-C\Big]\Big].
\end{eqnarray}

In Fig.(2), the graphs are drawn for the choice of the same parameters
given in Fig.(1), corresponding to m=1 . The throat of the wormhole is located
at r$\sim$1.5, where $\mathcal{G}$ cuts r-axis and the shape function
satisfy all the necessary conditions of as shown in the Figs.(2).

 Here also from Fig.(2), (lower left) the Null Energy Condition (NEC) and the Weak
Energy Condition (WEC) are violated, conditions that
are necessary to hold a wormhole open. It is interesting
to note that the Strong Energy Condition (SEC) is satisfied
 which shown in Fig(2), (lower right).

\begin{figure*}[thbp]
\begin{tabular}{rl}
\includegraphics[width=7.5cm]{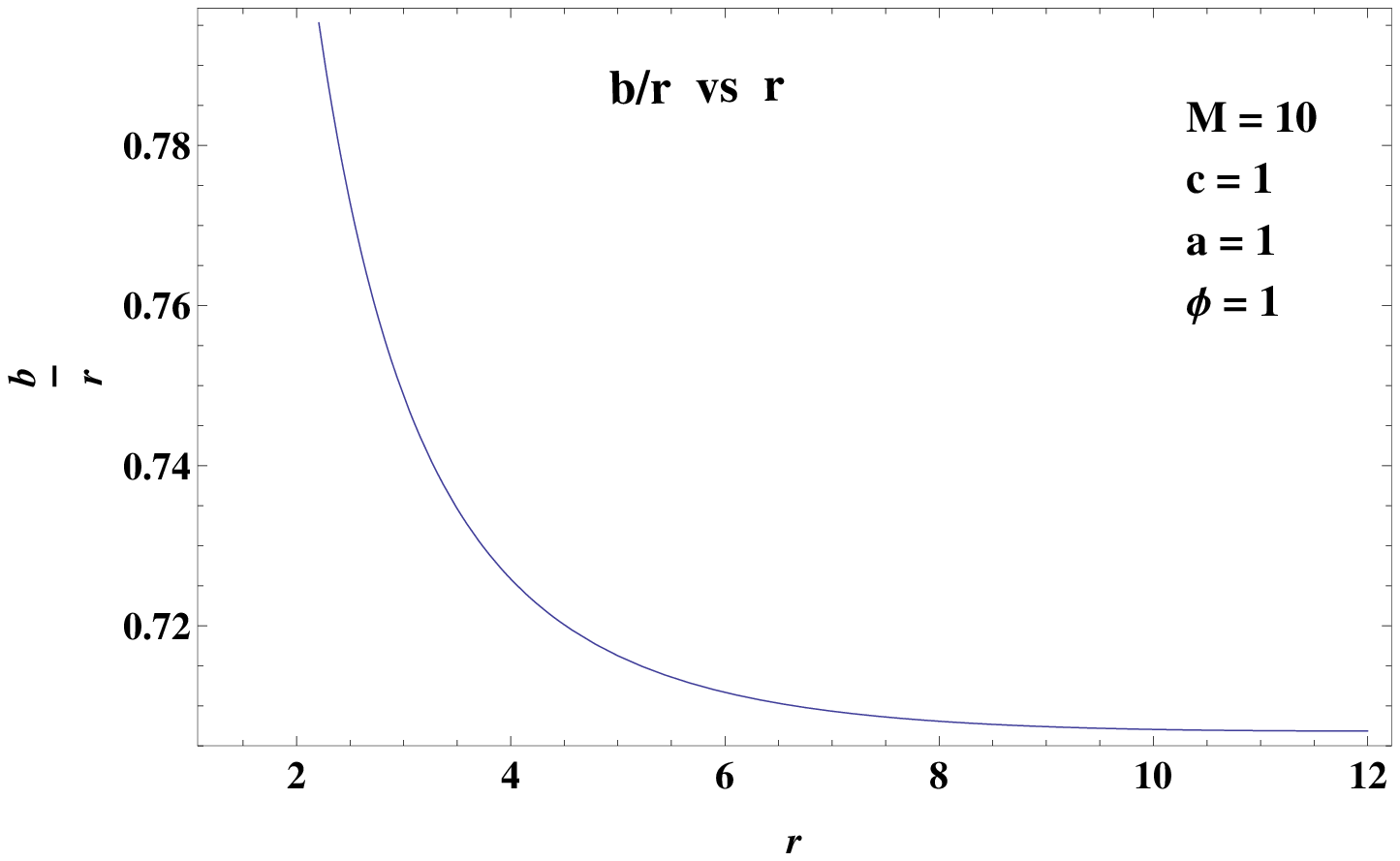}&
\includegraphics[width=7.5cm]{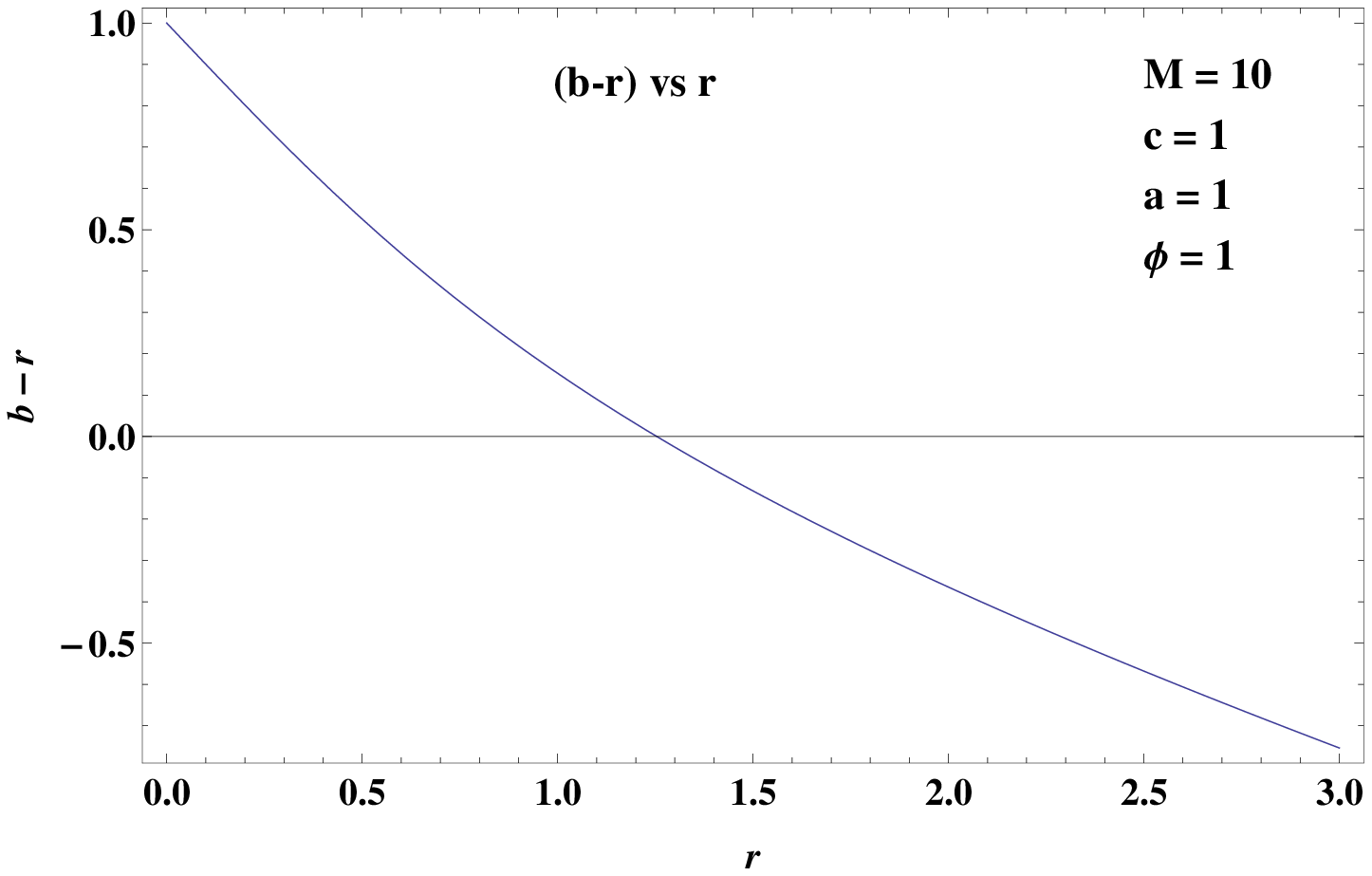} \\
\includegraphics[width=7cm]{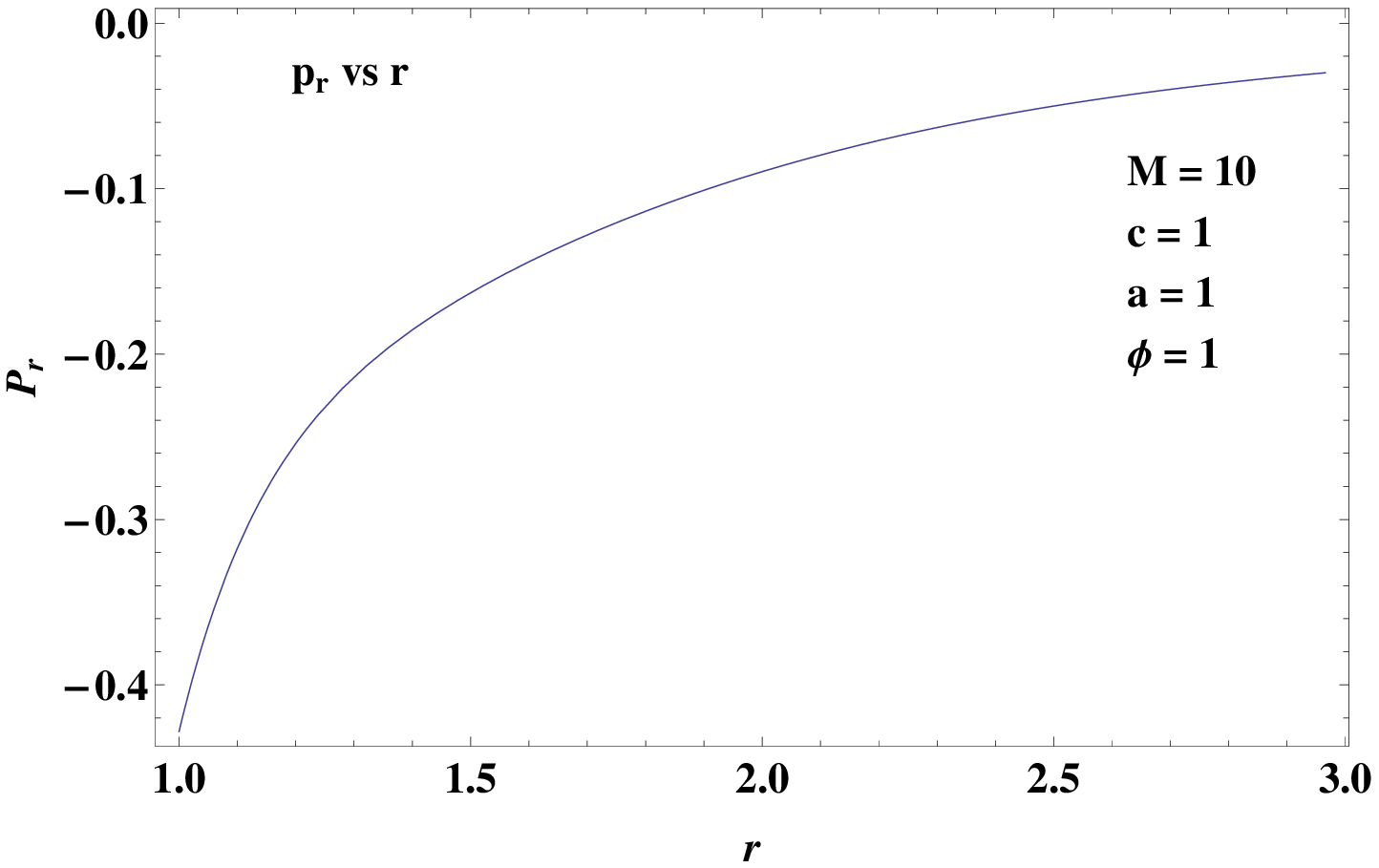}&
\includegraphics[width=7cm]{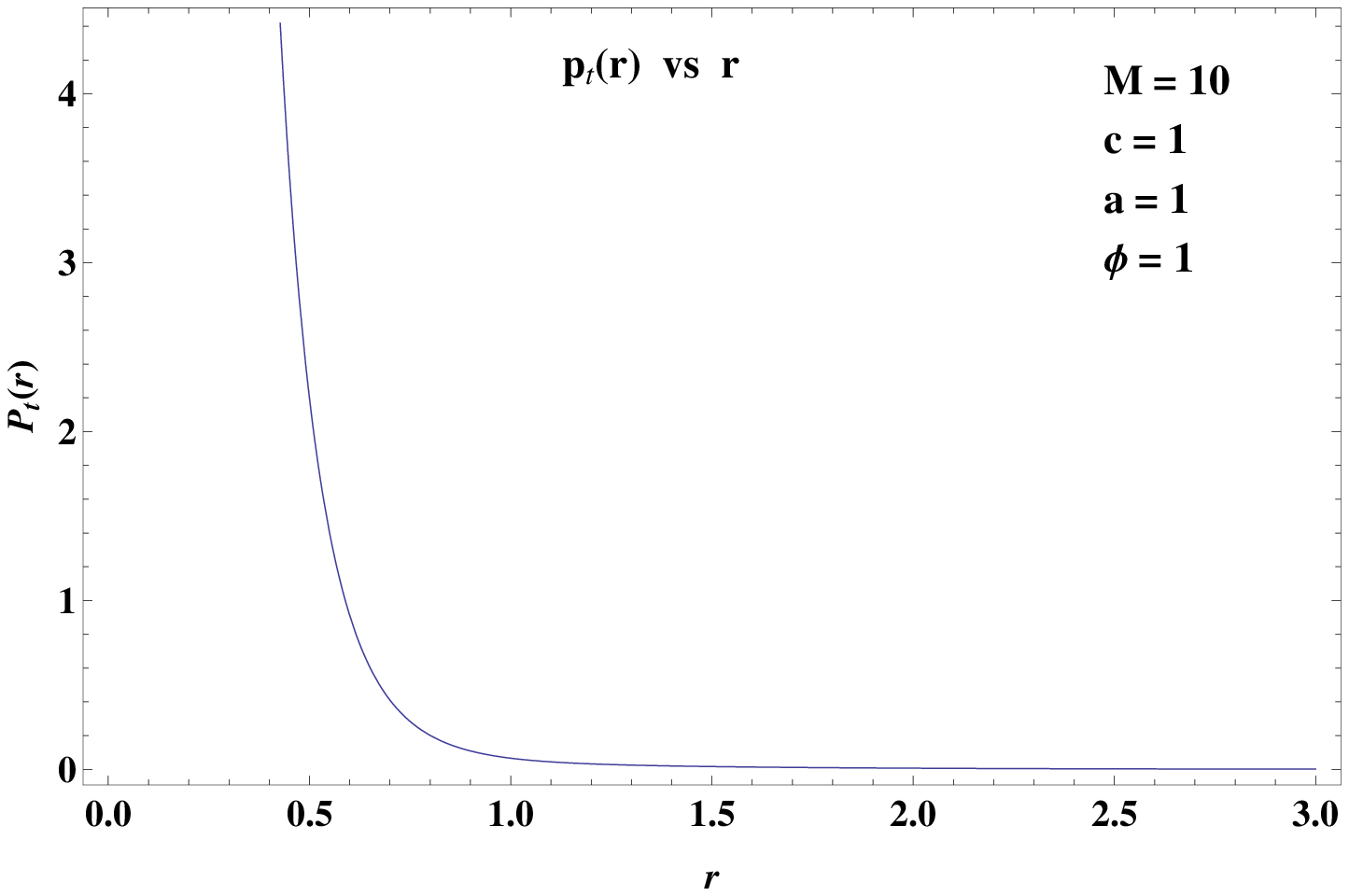} \\
\includegraphics[width=7cm]{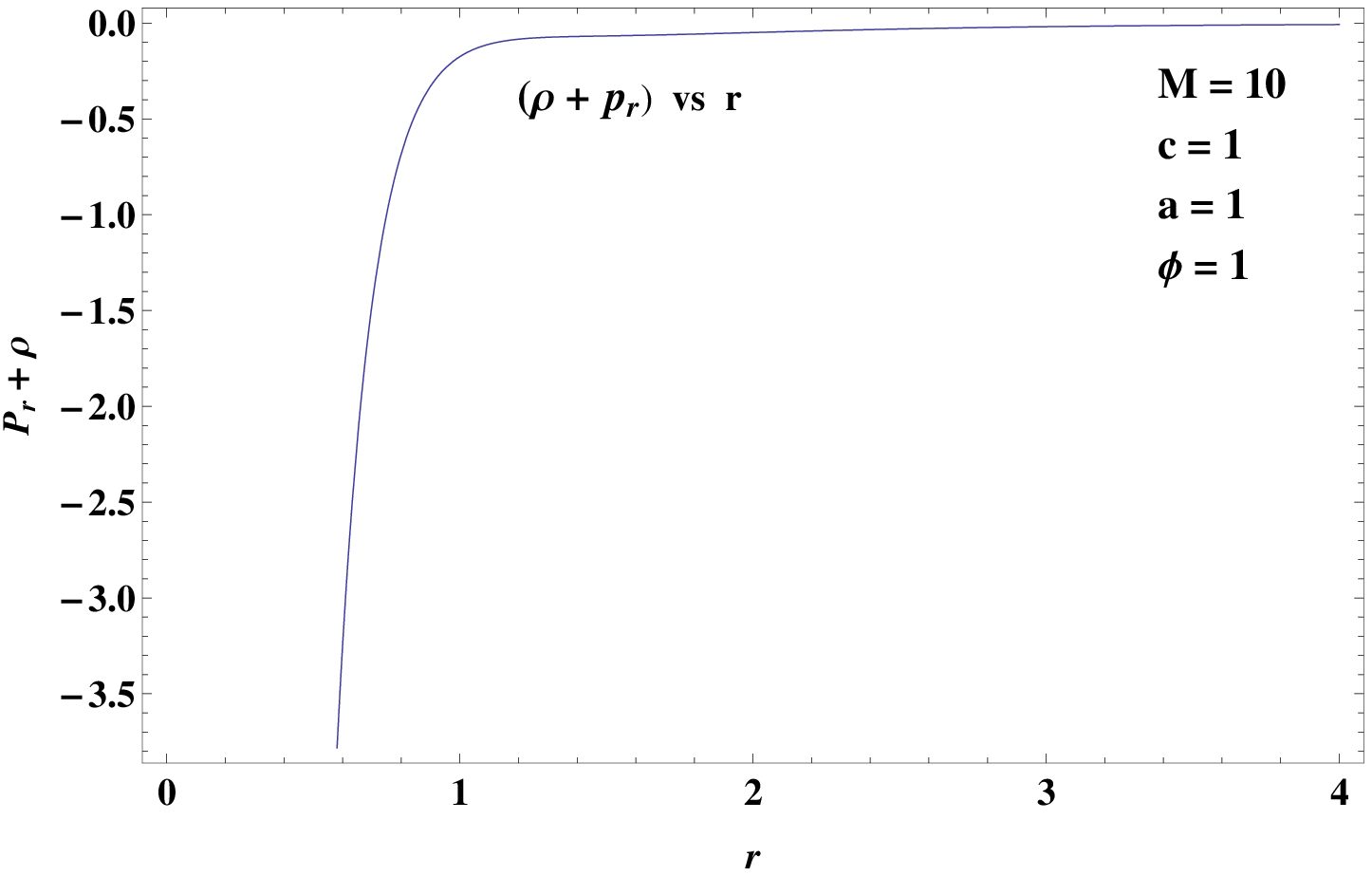} &
\includegraphics[width=7cm]{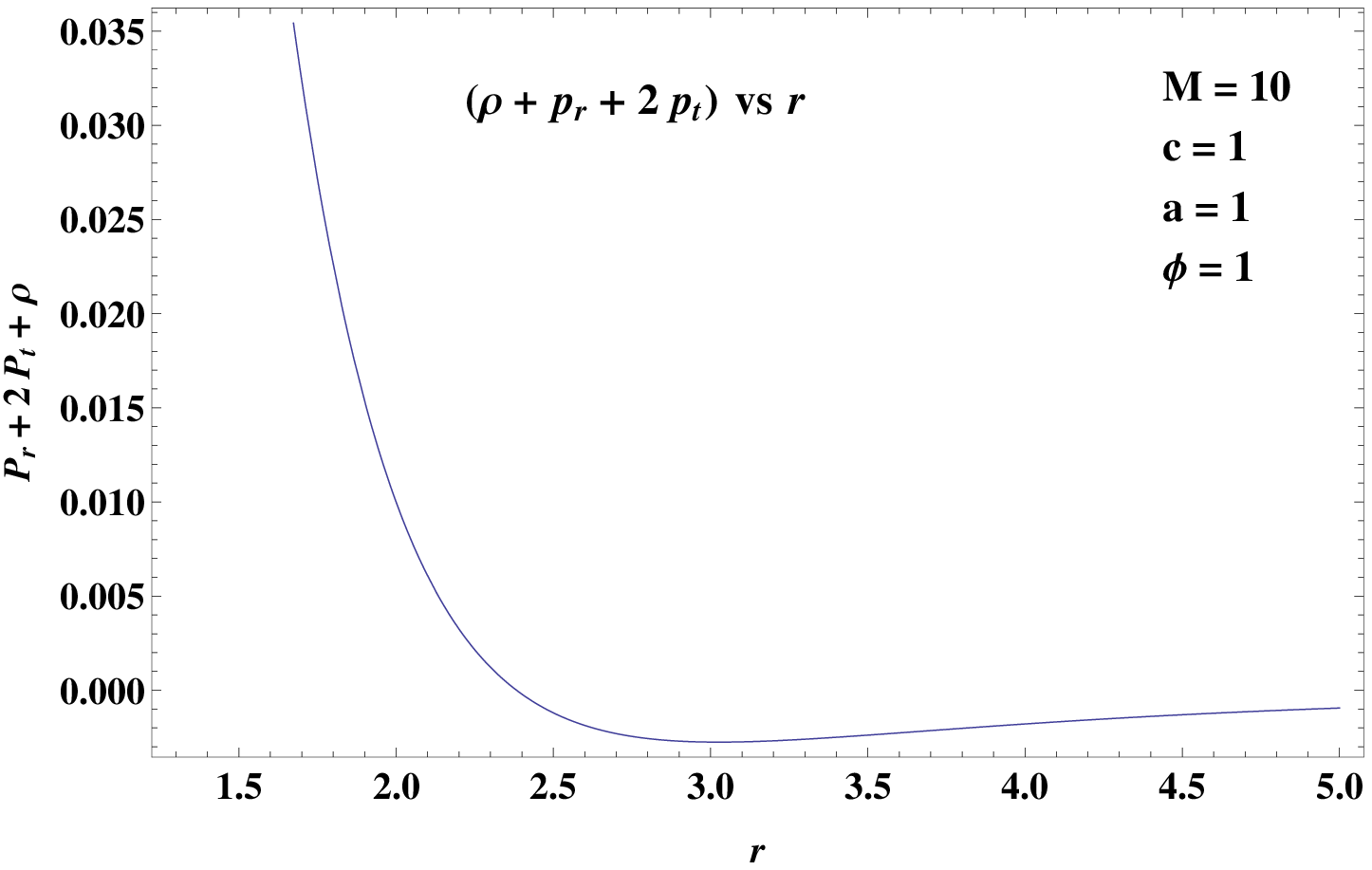} \\
\end{tabular}
\caption{ Plots for case $m=1$.}
\end{figure*}

\section{Wormhole solution for a given shape function}

We developed the literature by considering several interesting
shape function.
\subsection{$b(r)=r_0\Big(\frac{r}{r_0}\Big)^\alpha$}

Here, we consider a shape function of the form
\begin{equation}\label{br}
b(r)=r_0\Big(\frac{r}{r_0}\Big)^\alpha,
\end{equation}
which satisfy the flare-out conditions for $b^{\prime}(r_0)= \alpha < 1$,
and that for r $\rightarrow \infty$ we have b(r)/r = $(r_0/r)^{1-\alpha} \rightarrow 0$.

 Putting Eq. (19) in (6), we
have
\begin{equation}
R(r)=\frac{2\alpha}{r^2}\Big(\frac{r}{r_0}\Big)^{\alpha-1}.
\end{equation}
and, Substituting Eqs. (8) and
(19) in (3), we get
\begin{equation}
F(r)=\frac{M\sqrt{\phi}}{\alpha
\pi^2}\Big(\frac{r_0}{r}\Big)^{\alpha-1}\Big[\frac{r^{2}}{(r^2+\phi)^2}\Big].
\end{equation}
Using gravitational field equations (3)-(5),
the components of radial and transverse pressure, given by the following form
\begin{eqnarray}
p_r(r)&=&\frac{M\sqrt{\phi}}{\alpha \pi^2\Big(r^2+\phi\Big)^2}
\Big[-1+\frac{(\alpha-1)}{2}\Big\{(3-\alpha)-\frac{4r^2}{r^2+\phi}\Big\}+
\Big(1-\Big(\frac{r}{r_0}\Big)^{1-\alpha}\Big)\Big\{(3-\alpha)(2-\alpha)\nonumber\\&&
+\frac{4(2\alpha-5)r^2}{\Big(r^2+\phi\Big)}+\frac{8r(2r^3-\phi)}{\Big(r^2+\phi\Big)^2}\Big\}\Big],\\
p_t(r)&=&\frac{M\sqrt{\phi}}{\alpha \pi^2\Big(r^2+\phi\Big)^2}
\Big[\frac{(1-\alpha)}{2}+\Big(1-\Big(\frac{r}{r_0}\Big)^{1-\alpha}\Big)\Big\{(3-\alpha)-\frac{4r^2}{r^2+\phi}\Big\}
\Big].
\end{eqnarray}

In Fig.(3), the graphs are drawn for the choice of the parameters
M=10, $r_0 = 1$, $\alpha = -1$ and $\phi =1$, respectively. Here also the throat of the wormhole is located
at r$\sim$1.5, where $\mathcal{G}$ cuts r-axis shown in the Fig.(3), (upper right) and
the energy conditions are same as in previous cases shown in left and right lower Fig.(3).

\begin{figure*}[thbp]
\begin{tabular}{rl}
\includegraphics[width=7.5cm]{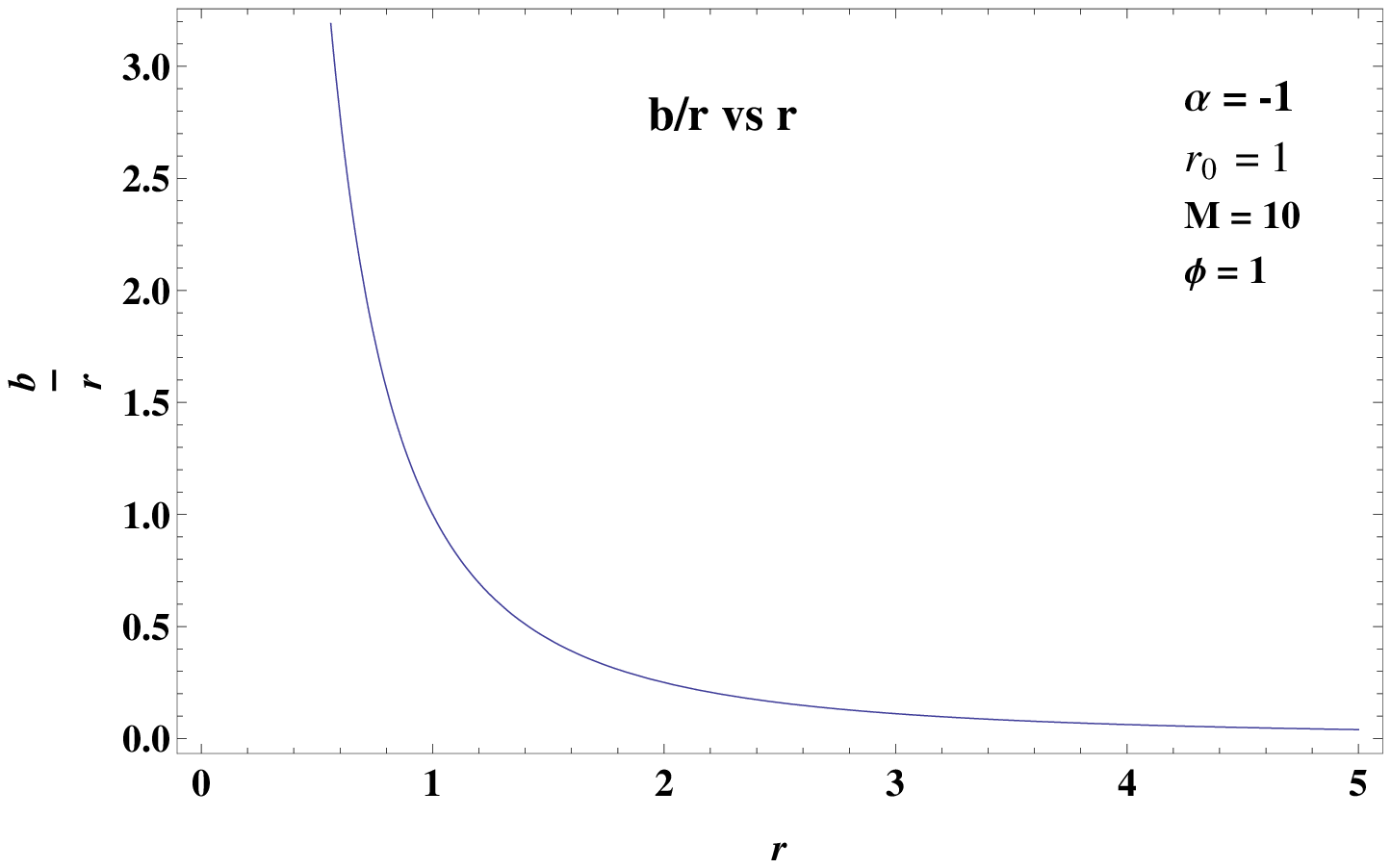}&
\includegraphics[width=7.5cm]{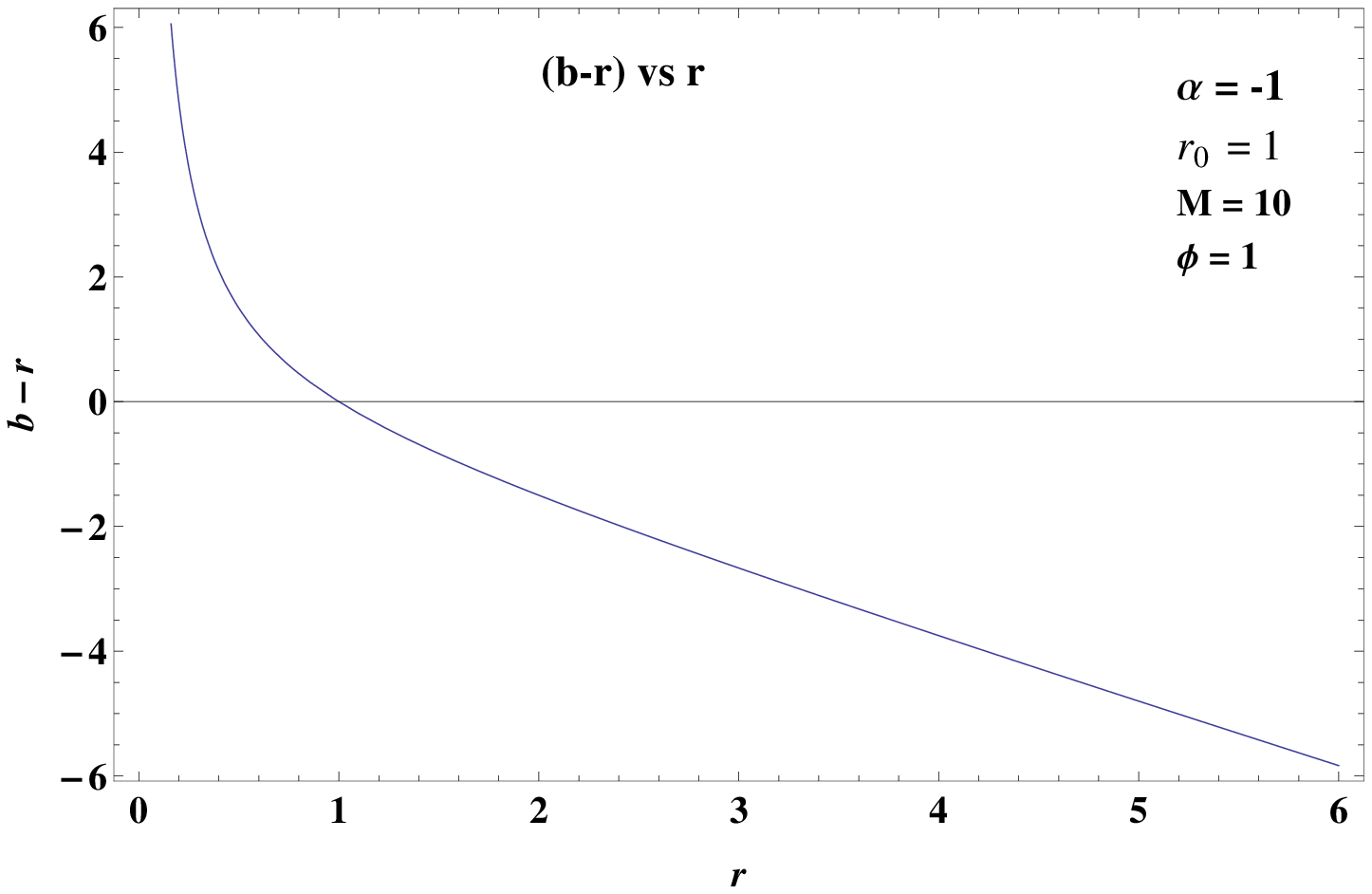} \\
\includegraphics[width=7cm]{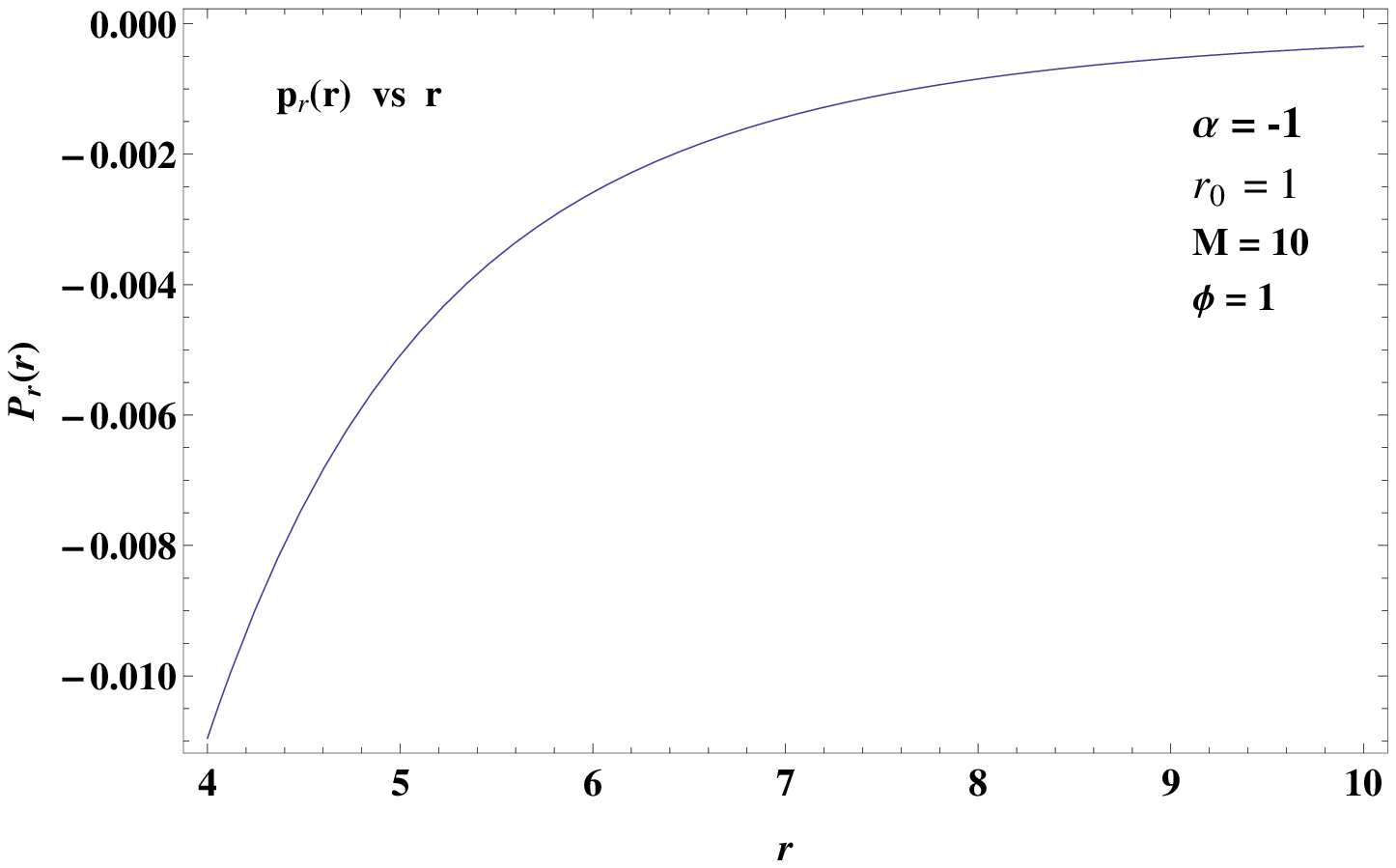}&
\includegraphics[width=7cm]{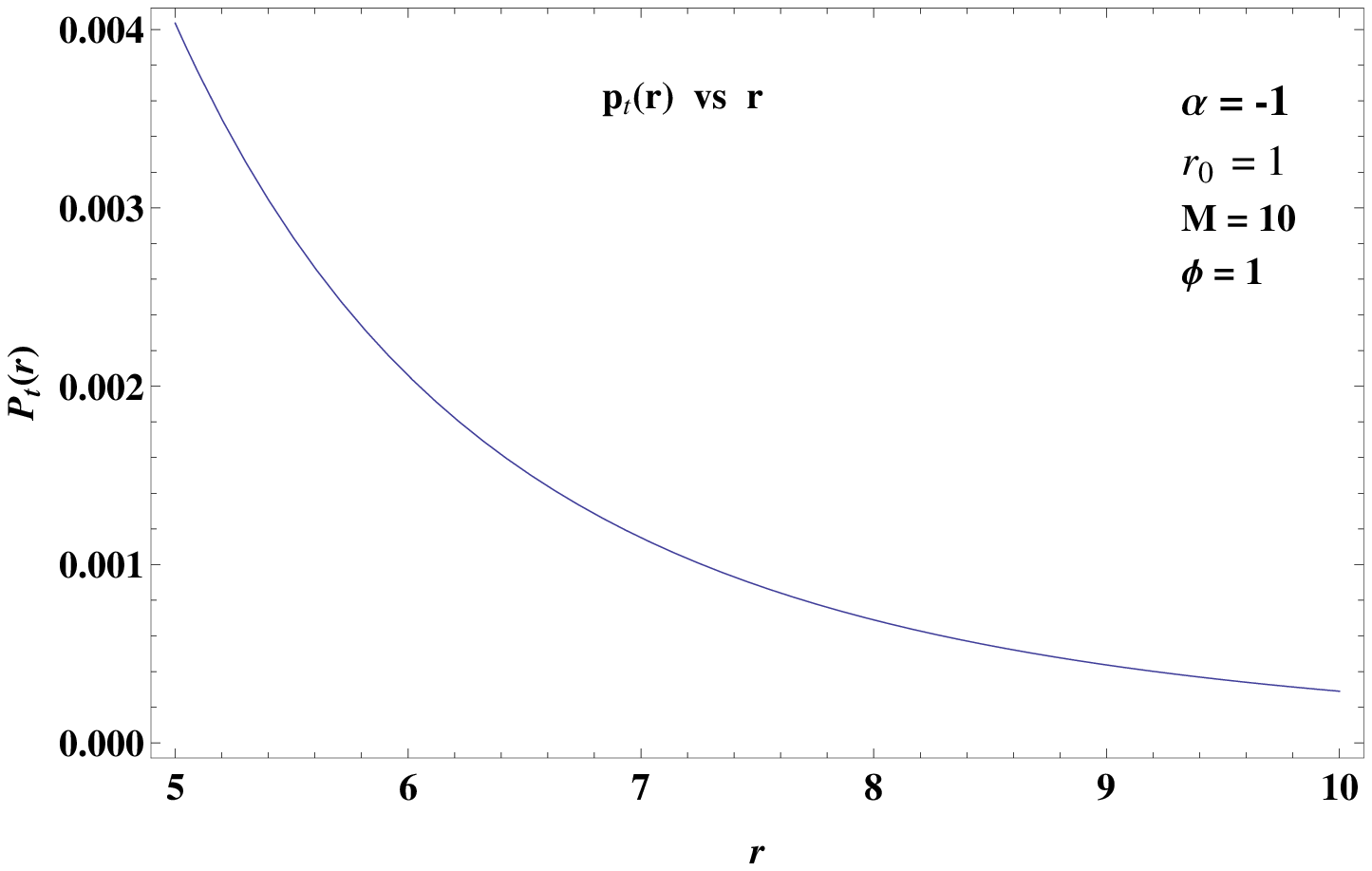} \\
\includegraphics[width=7cm]{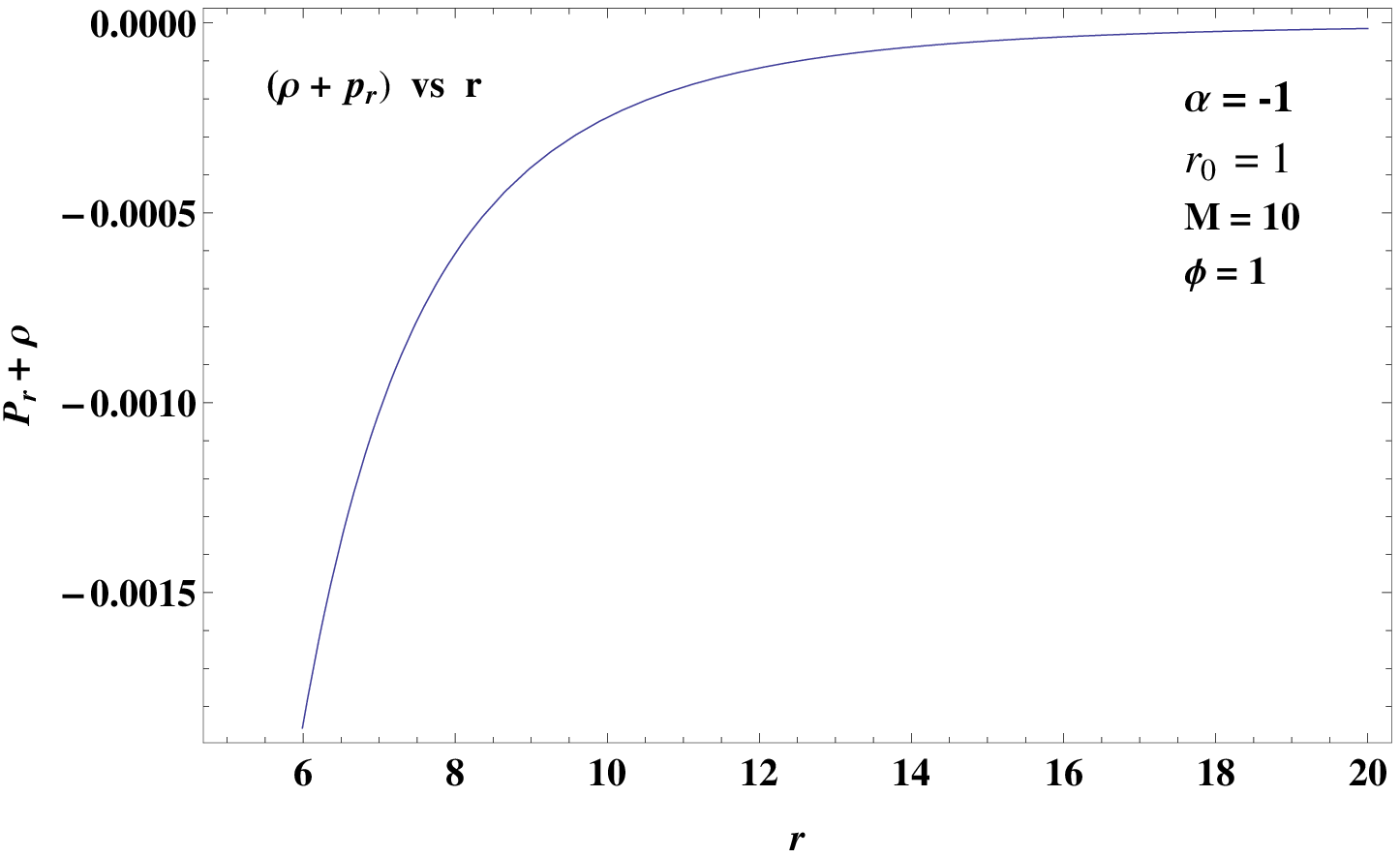} &
\includegraphics[width=7cm]{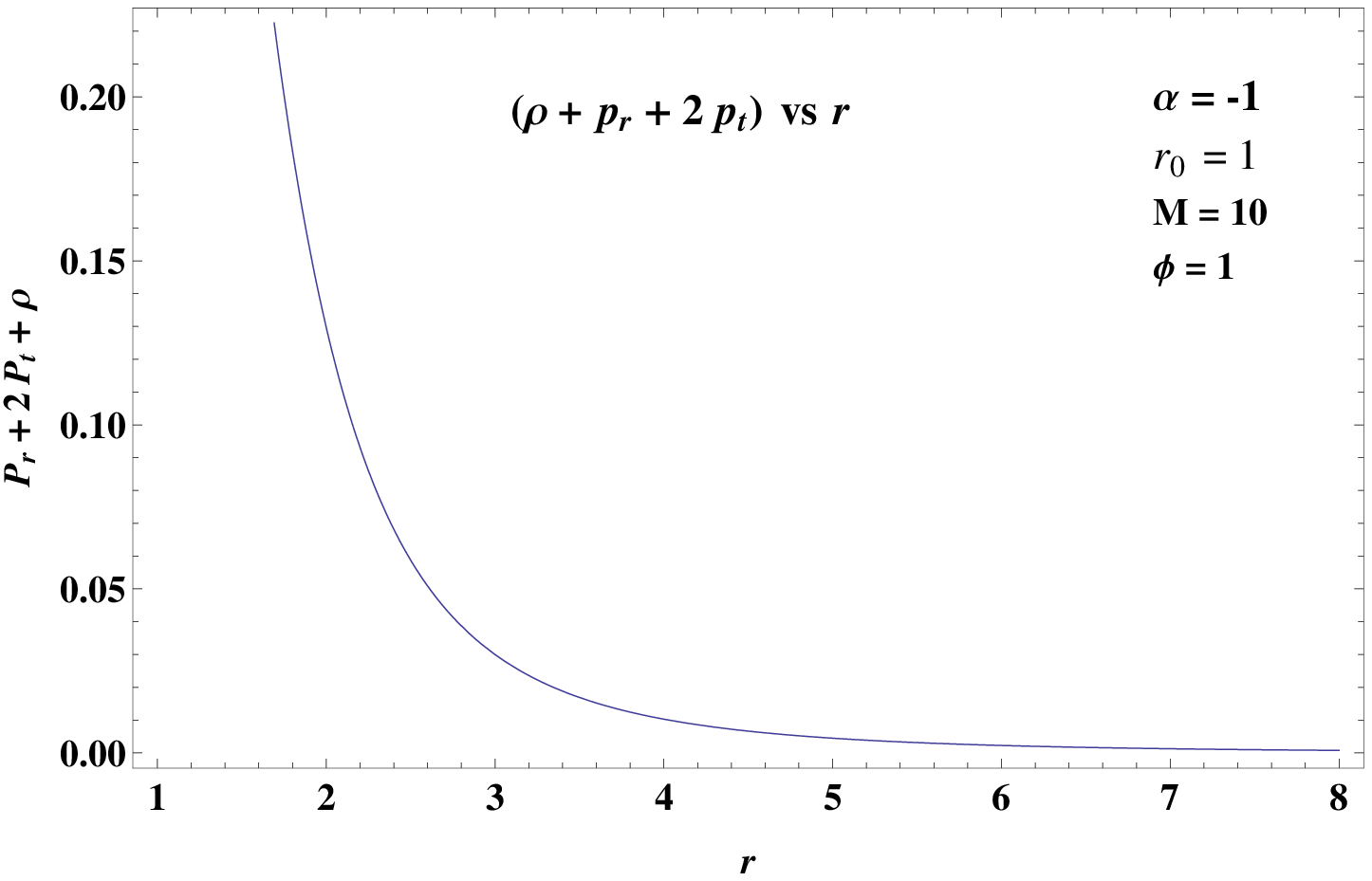} \\
\end{tabular}
\caption{ Plots for the case A of Section IV. }
\end{figure*}

\subsection{A particular shape function : $ b(r)=A tan^{-1}Cr$}

Let us consider a shape function of the form
\begin{equation}\label{br}
b(r)=A tan^{-1}Cr,
\end{equation}
so that b(r)/r = $A tan^{-1}Cr/r$ $\rightarrow 0$ for r $\rightarrow $ $\infty$
(by L'Hospital's rule),  which met the asymptotically flat conditions.
However, one
may also verify that $b^\prime(r_0) = AC/ (1+C^2r^2_0) < 1$ i.e.,
the fundamental property of a wormhole is that a flaring
out condition at the throat is met for the choice of the parameters.

 using Eq. (24) in (6), we have

\begin{equation}
R(r)=\frac{2}{r^2}\Big(\frac{AC}{1+C^2r^2}\Big).
\end{equation}
The form of $F(r)$ function becomes
\begin{equation}\label{br}
F(r)=\frac{M\sqrt{\phi}}{\pi^2
AC}\frac{r^2(1+C^2r^2)}{(r^2+\phi)^2}.
\end{equation}

The components of radial and transverse pressure become
\begin{eqnarray}
     P_r(r)&=&-\frac{M\sqrt{\phi}}{\pi^2 AC}\Big[ \frac{A}{r} tan^{-1}
     Cr\frac{(1+C^2r^2)}{(r^2+\phi)^2}-\Big(\frac{ACr}{1+C^2r^2}-Atan^{-1} Cr\Big)
     \Big(\frac{1+2C^2r^2}{r(r^2+\phi)^2}-\frac{2(r+C^2r^3)}{(r^2+\phi)^3}\Big) \nonumber\\&&   +\Big(1-\frac{A}{r}tan^{-1}
        Cr\Big)\Big(\frac{2+12C^2r^2}{(r^2+\phi)^2}-\frac{4(5r^2+9C^2r^4)}{(r^2+\phi)^3}+\frac{24r(r^3+C^2r^5)}{(r^2+\phi)^4}\Big)
        \Big],\\
P_t(r)&=&-\frac{M\sqrt{\phi}}{\pi^2 AC}
\Big[\Big(1-\frac{A}{r}tan^{-1}
     Cr\Big)\Big(\frac{2+4C^2r^2}{(r^2+\phi)^2}-\frac{4(r^2+C^2r^4)}{(r^2+\phi)^3}\Big)
     \nonumber\\&& +\Big(\frac{AC}{1+C^2r^2}-\frac{A}{r}tan^{-1} Cr\Big)\Big(\frac{1+C^2r^2}{2(r^2+\phi)^2}\Big)\Big]  .
\end{eqnarray}

In Fig.(4), the graphs are drawn for the choice of the parameters
M=10, a=1, A=2,  C=1 and $\phi =1$, respectively. Here also the
throat of the wormhole is located at r$\sim$ 2, where $\mathcal{G}$
cuts r-axis shown in the Fig.(4), (upper right) and the energy
conditions are same as in previous cases shown in left and right
 lower portion of Fig.(4).

\begin{figure*}[thbp]
\begin{tabular}{rl}
\includegraphics[width=7.5cm]{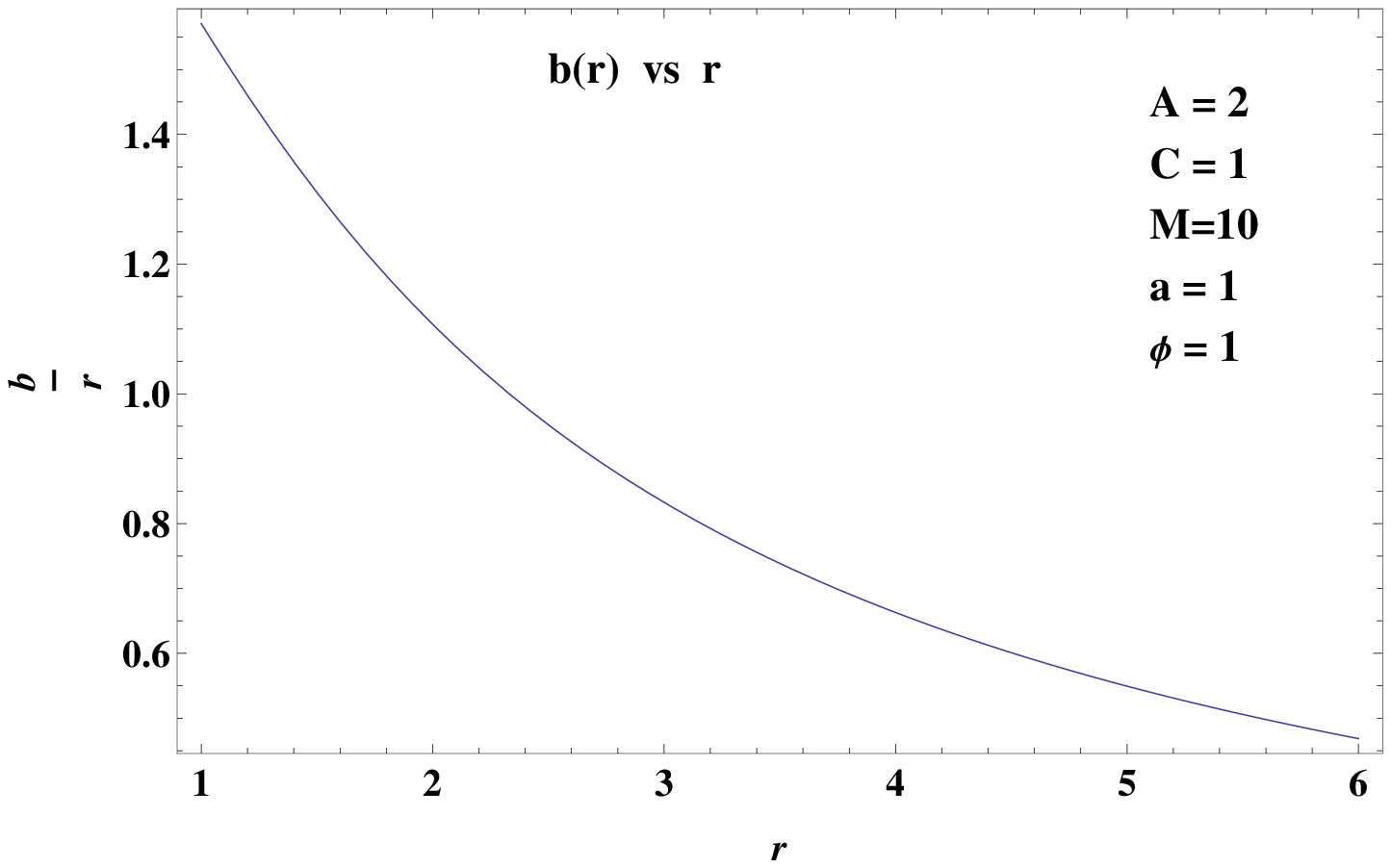}&
\includegraphics[width=7.5cm]{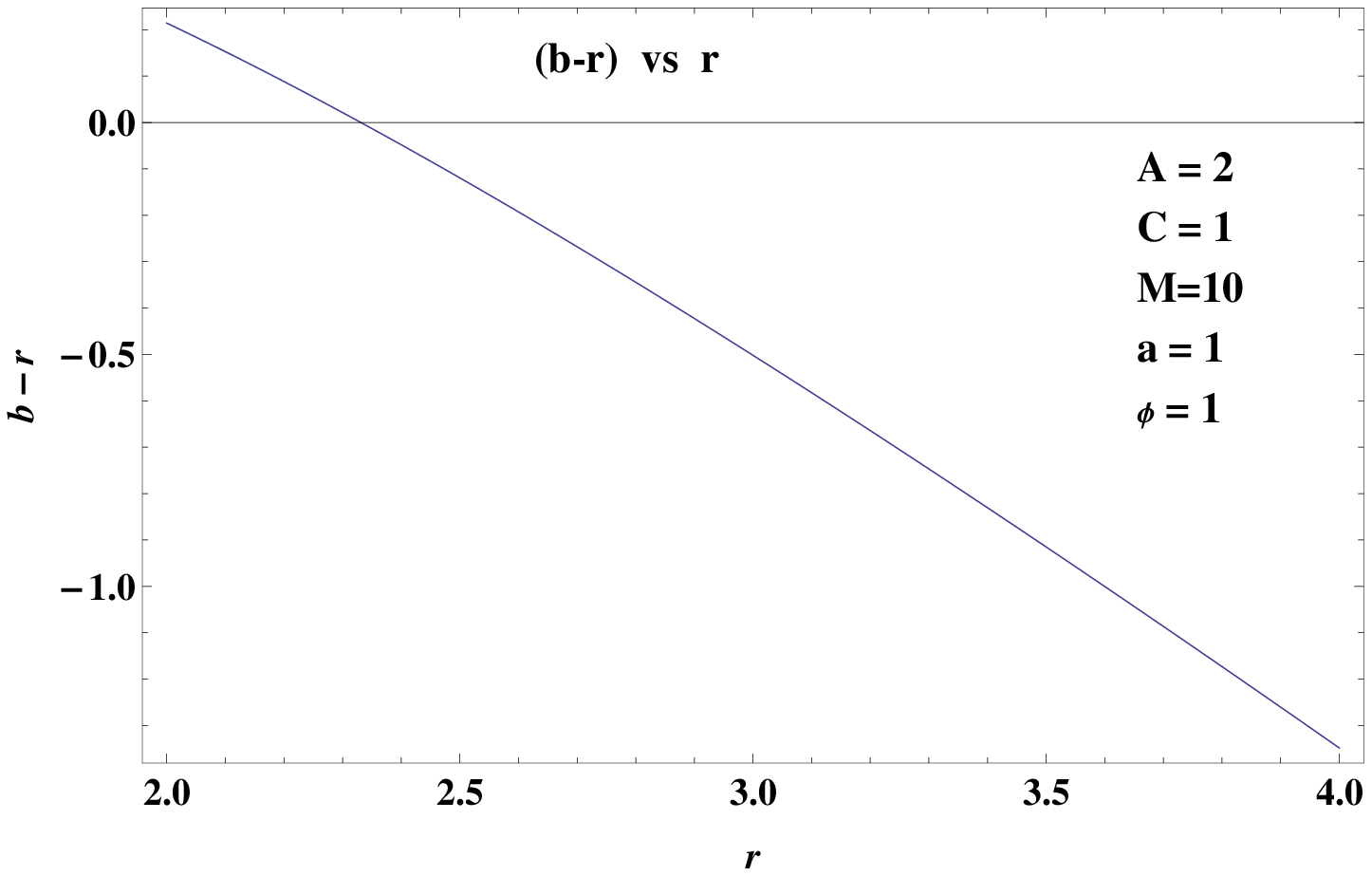} \\
\includegraphics[width=7cm]{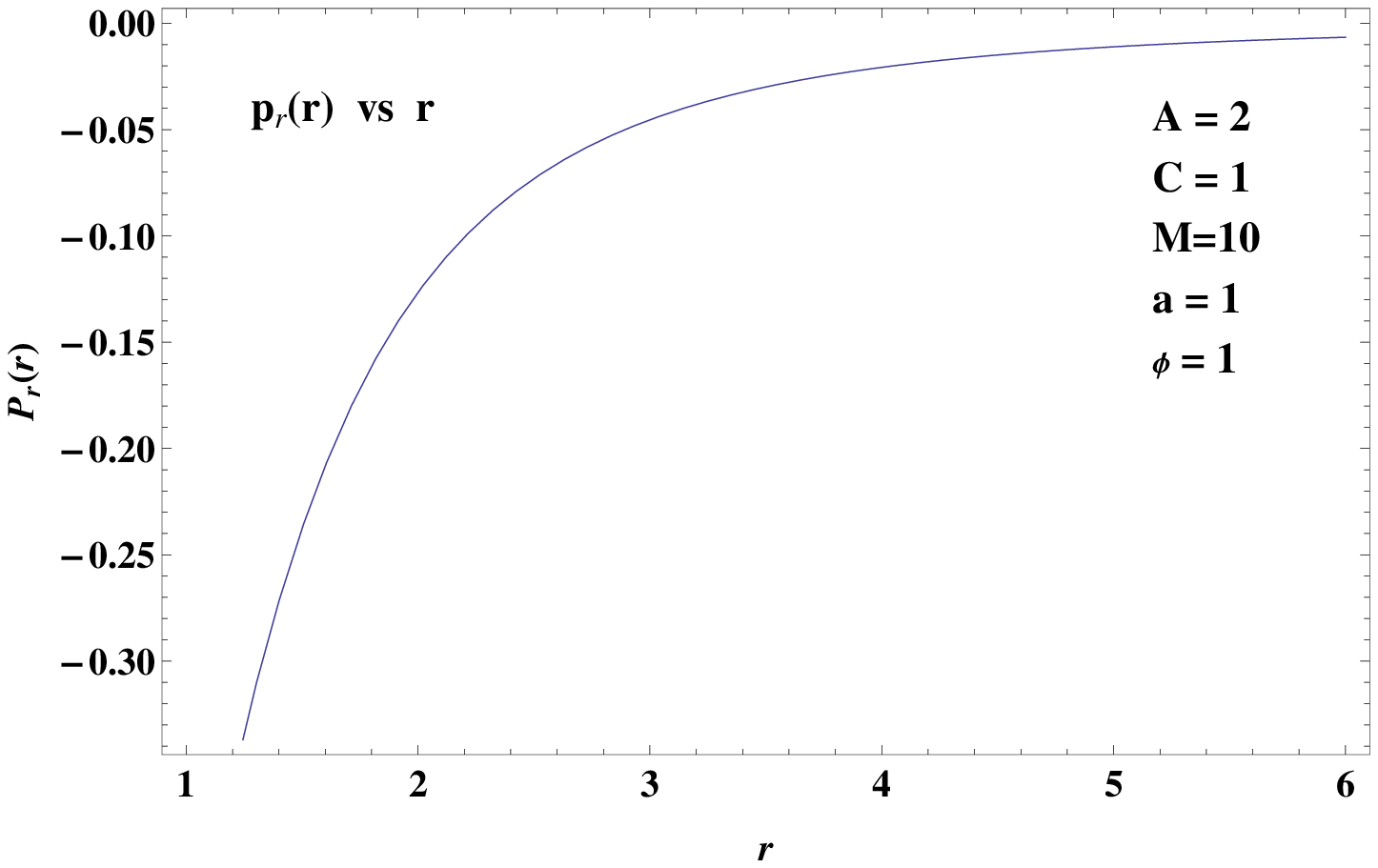}&
\includegraphics[width=7cm]{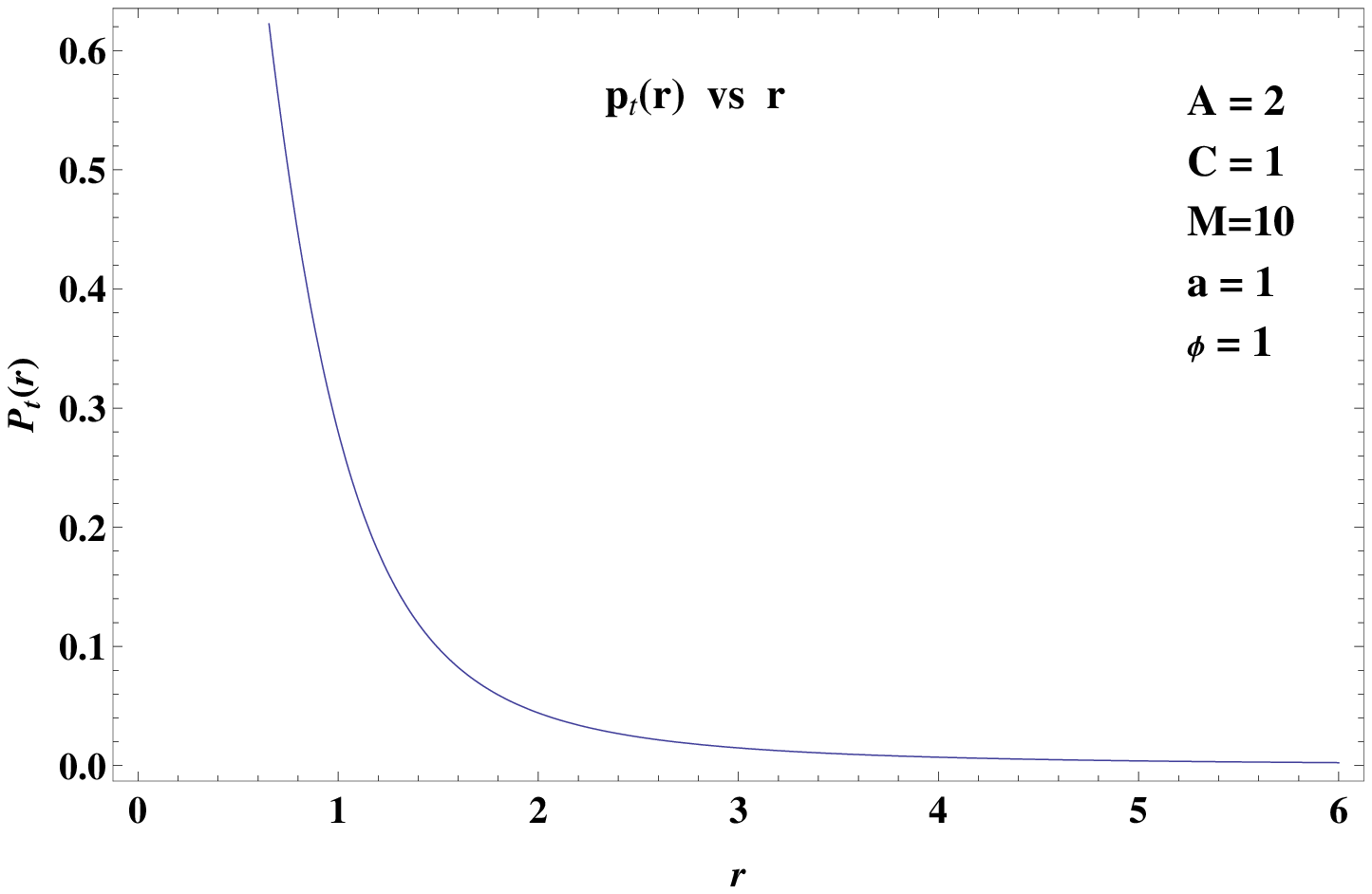} \\
\includegraphics[width=7cm]{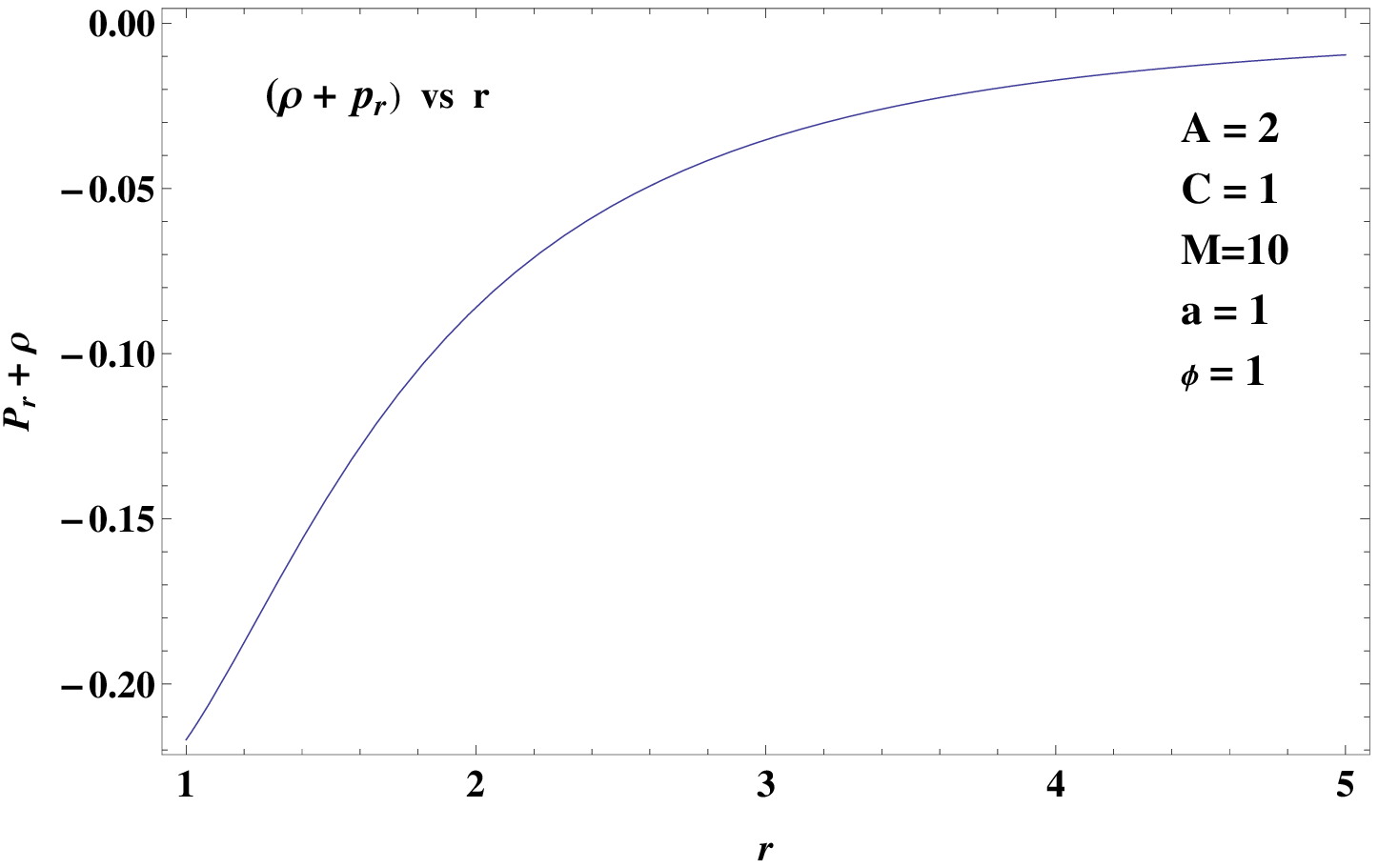} &
\includegraphics[width=7cm]{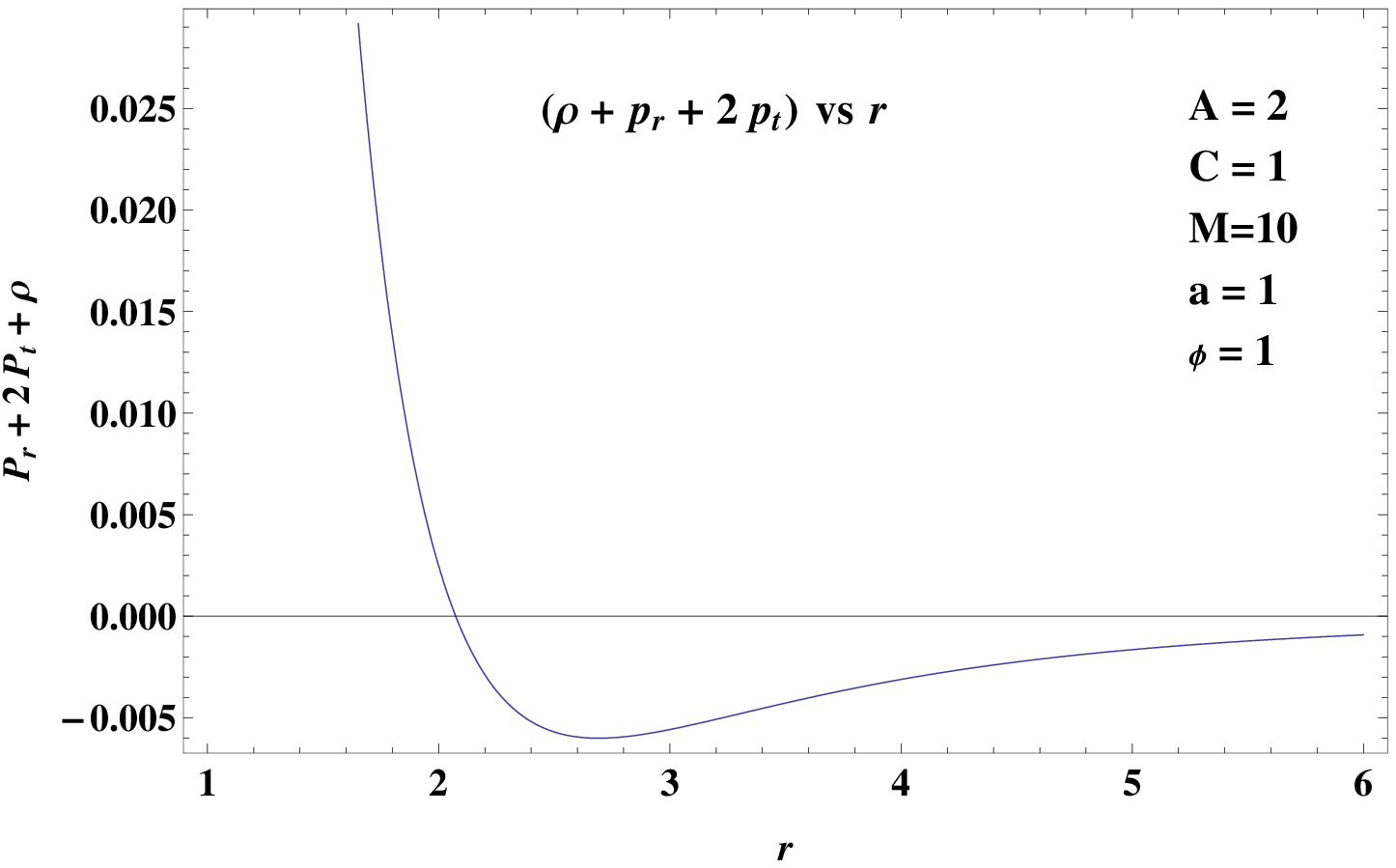} \\
\end{tabular}
\caption{ Plots of case B of Section IV. }
\end{figure*}

\pagebreak

\section{Concluding remarks}

In this paper, we derived some new exact solutions of static
wormholes in $f(R)$ gravity supported by the matter possesses
Lorentizian density distribution of a  particle-like  gravitational
source. We derive the wormhole's solutions in two possible schemes
for a given Lorentzian distribution. The first model assumes the
power-law form F(R) whereas the second model discussed assumes a
particular shape function which allows the reconstruction of f(R).
For the power-law form  of F(R) with m=1 is interesting as in this
case  the null energy condition is
  violated, but the strong energy condition is met.
For the second model, we have considered two  particular shape
functions and have   reconstructed   f(R) in both cases. In these
two cases, the null energy condition are once again violated, but
the strong energy conditions are  met. All the solutions assume zero
tidal forces which automatically that the wormholes are traversable
\cite{lobo1}.

\pagebreak

\section*{Acknowledgments} FR   is thankful to the authority of
Inter-University Centre for Astronomy and Astrophysics, Pune,  India
for providing them Visiting Associateship under which a part of this
work was carried out. AB is also thankful to IUCAA for giving him an
opportunity to visit IUCAA where a part of this work was carried
out. FR is also thankful   UGC, Govt. of India under research award
scheme,  for providing financial support.

\end{document}